\documentclass[aps,prd,preprintnumbers,groupedaddress,nofootinbib,amssymb,notitlepage,eqsecnum]{revtex4-2}
\usepackage{
here}
\usepackage{iftex}

\ifPDFTeX
  \usepackage{graphicx}
\else
  \usepackage[dvipdfmx]{graphicx}
\fi
\usepackage{amsmath,amsthm,amssymb}
\usepackage{bm}
\usepackage{color}

\usepackage{amsfonts}
\usepackage{dcolumn}
\usepackage{hyperref}
\allowdisplaybreaks[1]
\usepackage{stackengine}

\usepackage{lmodern}
\usepackage{microtype}
\usepackage{physics}
\usepackage{enumitem}

\newcommand{\be}{\begin{equation}}  
\newcommand{\ee}{\end{equation}}
\newcommand{\ba}{\begin{eqnarray}}
\newcommand{\ea}{\end{eqnarray}}

\newcommand{\rd}{{\rm d}}

\newcommand{\bem}{\begin{bmatrix}}
\newcommand{\eem}{\end{bmatrix}}
\newcommand{\Mpl}{M_{\rm Pl}}

\allowdisplaybreaks

\begin{document}

\preprint{YITP-26-54, WUCG-26-04}

\title{Inspiral gravitational waveforms from charged compact binaries with scalar hair}

\author{Antonio De Felice$^{a}$}
\email{antonio.defelice@yukawa.kyoto-u.ac.jp}  

\author{Shinji Tsujikawa$^{b}$}
\email{tsujikawa@waseda.jp} 

\affiliation{$^{a}$Center for Gravitational Physics 
and Quantum Information, Yukawa Institute for 
Theoretical Physics, Kyoto University, 
606-8502, Kyoto, Japan\\
$^{b}$Department of Physics, Waseda University, 
3-4-1 Okubo, Shinjuku, Tokyo 169-8555, Japan}

\begin{abstract}

We investigate gravitational waveforms from compact binary systems in Einstein-scalar-Maxwell (ESM) theories, where a scalar field $\phi$ couples to a $U(1)$ gauge field $A_\mu$ through a field-dependent function $\mu(\phi)$. In this framework, compact objects---black holes (BHs), neutron stars (NSs), and exotic compact objects (ECOs)---can carry both vector and scalar charges, with the latter arising as secondary hair induced by the former. Modeling the binary as electrically charged point particles with scalar-field-dependent masses, we derive the conservative dynamics in the near zone and compute the radiative fields in the far zone. The tensor waveform is modified through the effective dynamics and radiation-reaction-driven phase evolution, while scalar and vector modes introduce additional energy-loss channels. From the energy fluxes of tensor, scalar, and vector radiation, we construct the frequency-domain waveform using the stationary phase approximation. Dipole radiation sourced by differences in scalar and vector charge-to-mass ratios yields a leading 
$-1$~post-Newtonian correction. The deviation from general relativity is characterized by a single parameter $b$, which controls both amplitude and phase modifications. We further examine constraints from the orbital-period decay of binary pulsars, showing that current observations already place stringent bounds on $b$ for neutron star binaries. 
In addition, we evaluate $b$ for representative
BH-BH, NS-NS, ECO-ECO binaries realized in ESM theories. 
Our results provide a unified framework for gravitational-wave signatures of
charged compact binaries and offer a means of testing dark-sector
scalar and vector charges with current and future observations.
\end{abstract}

\date{\today}


\maketitle

\tableofcontents

\section{Introduction}

The direct detection of gravitational waves (GWs) by the LIGO-Virgo-KAGRA collaboration 
has opened a new avenue for probing gravity in the strong-field regime. 
Following the first observation of a binary black hole (BH) merger \cite{LIGOScientific:2016aoc}, numerous compact binary 
coalescences---including neutron star (NS) systems---have been detected. 
In particular, the NS-NS merger event GW170817 \cite{LIGOScientific:2017vwq}, accompanied by an electromagnetic counterpart \cite{Goldstein:2017mmi}, marked the beginning of multimessenger astrophysics \cite{LIGOScientific:2017ync,LIGOScientific:2017zic}.
This event enabled the measurement of tidal deformability in NSs, providing 
a powerful probe of the equation of state of dense 
matter \cite{LIGOScientific:2018cki}. Moreover, the speed of 
gravity has been constrained to be extremely close to that of light, thereby 
placing tight restrictions on dark energy 
models \cite{Creminelli:2017sry,Ezquiaga:2017ekz,Sakstein:2017xjx,Baker:2017hug,Kase:2018aps}.

Subsequent GW detections, such as the BH-NS merger 
GW200115 \cite{Abbott:2021qlt}, have further enriched 
both the observational and theoretical landscape. 
At the same time, some events have exhibited features that may hint at the presence of exotic compact objects (ECOs) beyond standard BHs and NSs. For instance, the event 
GW190521 \cite{Abbott:2020tfl}, characterized by its short duration and unusual waveform morphology, has been interpreted as a possible merger involving nonstandard compact objects such as boson stars. In addition, events like GW190814 \cite{Abbott:2020khf} and 
GW230529 \cite{Abbott:2024lxy}, which involve compact objects with masses in the so-called mass gap between NSs and BHs, may also point toward the existence of ECOs.
Overall, the observed signals from compact binary coalescences have been broadly consistent with the predictions of general relativity (GR) with standard matter. Nevertheless, they provide an unprecedented opportunity to search for deviations from GR and to probe additional degrees of freedom, such as scalar and vector fields in the dark sector.

In the case of a scalar field $\phi$, the existence of hairy BH solutions 
with a radially dependent profile $\phi(r)$ on a static, spherically symmetric background is highly constrained~\cite{Hui:2012qt,Minamitsuji:2022mlv,Minamitsuji:2022vbi}, even within the framework of the most general scalar-tensor theories with second-order equations of motion (Horndeski theories~\cite{Horndeski:1974wa}).
Indeed, it has been shown that BHs do not acquire scalar hair in a variety of theories, 
including minimally coupled canonical scalar fields~\cite{Hawking:1972qk,Bekenstein:1972tm}, 
k-essence~\cite{Graham:2014ina}, nonminimally coupled scalars of 
the form 
${\cal F}(\phi)R$~\cite{Hawking:1972qk2,Bekenstein:1995un,Sotiriou:2011dz,Faraoni:2017xlh}, 
and Galileon theories~\cite{Hui:2012qt}.
A notable exception arises in theories where the scalar field 
couples to the Gauss-Bonnet (GB) invariant $R_{\rm GB}^2$ 
via a term of the form $\xi(\phi) R_{\rm GB}^2$ \cite{Kanti:1995vq,Torii:1996yi,Kanti:1997br,Pani:2009wy,Sotiriou:2013qea,Doneva:2017bvd,Silva:2017uqg,Antoniou:2017acq}.
However, in scalar-GB theories, the propagation speed of GWs 
generally deviates from that of light \cite{Kobayashi:2011nu,Kobayashi:2012kh,Kobayashi:2014wsa,Kase:2021mix}. 
Therefore, hairy BH solutions in such theories must 
be consistent with 
the stringent constraints from GW measurements. 
Moreover, scalar-GB theories that exhibit 
spontaneous scalarization of BHs can also face nontrivial cosmological constraints: the same coupling
may induce instabilities on cosmological backgrounds, or require a tuning of the initial scalar-field value so that 
the scalar remains dormant during
the cosmological evolution \cite{Anson:2019uto,Franchini:2019npi}.

For NSs, the presence of matter inside the star leads to a nonvanishing 
Ricci scalar $R$, which is related to the trace of the energy-momentum 
tensor $T$. If a scalar field $\phi$ is nonminimally coupled to $R$ 
through a term of the form ${\cal F}(\phi)R$, the NS can develop nontrivial 
scalar hair through this gravity-mediated interaction. Hairy NS solutions 
carrying a scalar charge can arise, for instance, in scalar-tensor theories 
of the Brans-Dicke type \cite{Brans:1961sx}, with a coupling function such as 
${\cal F}(\phi)=e^{-2Q\phi/\Mpl}$, 
where $\Mpl$ is the reduced 
Planck mass and $Q$ is a dimensionless constant. For coupling functions ${\cal F}(\phi)$ with an even power-law dependence on $\phi$, e.g., 
${\cal F}(\phi)=e^{-\beta \phi^2/(2\Mpl^2)}$, it is well known that 
spontaneous scalarization of NSs can 
occur \cite{Damour:1993hw}.
In this case, a tachyonic instability in the strong-gravity regime can trigger 
a transition from the GR branch $\phi=0$ to a scalarized branch $\phi \neq 0$.
In such theories, NSs acquire a scalar charge $q_s$, whereas BHs do not.
Gravitational waveforms emitted during the inspiral phase of compact 
binaries \cite{Eardley:1975js,Will:1994fb,Yunes:2009ke,Yunes:2009ef,Alsing:2011er,Yagi:2011xp,Berti:2012bp,Yagi:2012vf,Lang:2013fna,Mirshekari:2013vb,Lang:2014osa,Yagi:2016jml,Sennett:2016rwa,Sagunski:2017nzb,Bernard:2018ibn,Tahura:2018xdr,Liu:2020cpc,Bernard:2022ksr,Higashino:2022cnd,Quartin:2023tpl} provide a powerful probe of this effect; in particular, observations of the BH-NS merger event GW200115  \cite{Abbott:2021qlt} 
have been used to place constraints on the magnitude of 
the NS scalar charge \cite{Niu:2021nic,Takeda:2023wqn}.
 
For a $U(1)$ gauge-invariant vector field $A_\mu$, static and spherically symmetric BHs are described by the Reissner-Nordstr\"om (RN) solution \cite{Reissner:1916}, specified by the BH mass and its electric or magnetic charge. In the case of an electrically charged BH, however, the presence of surrounding plasma tends to neutralize the charge, as oppositely charged particles can be efficiently accreted onto the BH \cite{Bardeen:1970,Lee:1972}. This argument does not necessarily apply if the vector field belongs to a dark sector and is decoupled from Standard Model (SM) particles. In such a case, even if the BH carries a dark electric charge, the absence of interactions with the ambient plasma prevents efficient neutralization. Interestingly, the existence of such dark vector radiation is also motivated by cosmological observations, which allow for (and in some analyses mildly favor) extra relativistic degrees of freedom beyond the SM, commonly parameterized by the effective number of neutrino species $N_{\rm eff}$~\cite{Planck:2018vyg,Aghanim:2018eyx}. 

In the dark sector, one can also introduce a scalar field $\phi$ coupled to a $U(1)$ gauge field $A_{\mu}$. Interactions of this type arise generically in supergravity and string-inspired effective field theories, where scalar
fields such as dilatons and moduli couple nontrivially to gauge-field kinetic terms. 
In particular, the gauge kinetic function can be promoted
to a scalar-field-dependent function, leading to scalar-vector couplings of the form $\mu(\phi)F$, where $F=-F_{\mu\nu}F^{\mu\nu}/4$ and
$F_{\mu \nu}=\nabla_{\mu} A_{\nu}-\nabla_{\nu} A_{\mu}$, with
$\nabla_{\mu}$ denoting the covariant derivative
\cite{Gaillard:1981rj,Duff:1995sm,Andrianopoli:1996cm}.
Such couplings also arise naturally from higher-dimensional theories after dimensional reduction to four dimensions, where scalar fields
associated with compactification can couple to the effective gauge
kinetic term~\cite{Kaluza:1921tu,Klein:1926tv}. These theories belong
to the class of Einstein-scalar-Maxwell (ESM) theories, 
for which a representative four-dimensional Lagrangian 
density is given by
${\cal L}=\Mpl^2 R/2-(\nabla_\mu \phi)(\nabla^\mu \phi)/2+\mu(\phi)F$.

A representative example of a static, spherically symmetric,
electrically charged BH in ESM theories is provided by the
Gibbons-Maeda solution~\cite{Gibbons:1987ps} and the
Garfinkle-Horowitz-Strominger solution~\cite{Garfinkle:1990qj}.
It is obtained for a dilatonic coupling of the form
$\mu(\phi)=e^{-\phi}$ in units where $\Mpl=\sqrt{2}$.
Similar configurations also exist for more general 
exponential couplings, $\mu(\phi)=e^{-\gamma \phi/\Mpl}$, where $\gamma$ is a nonzero dimensionless constant.
In this class of solutions, the scalar field develops a nontrivial radial profile sourced by the electric charge $q$ of the BH. Importantly, the associated scalar charge $q_s$ is not an independent parameter, but is instead determined by $q$ and the BH mass. In this sense, the scalar hair constitutes secondary hair, induced by the presence of primary hair, 
namely the electric charge. If one considers a coupling function $\mu(\phi)$ that depends on an even power of $\phi$, 
spontaneous scalarization can also occur for electrically 
charged BHs~\cite{Herdeiro:2018qqt,Fernandes:2019rez,Ikeda:2019qjp}. In this case, the RN solution with sufficiently 
large electric charge becomes unstable, giving rise to scalarized BHs endowed with both 
scalar and vector hair.

In ESM theories, relativistic stars such as NSs can also admit electrically charged solutions with scalar hair.
Unlike BHs, NSs contain matter fluids in their 
interior, subject to regular boundary 
conditions at the stellar center. 
In particular, it was shown in Ref.~\cite{Minamitsuji:2021vdb} that spontaneously scalarized NS solutions arise for a scalar-vector coupling of the form 
$\mu(\phi)=e^{-\beta \phi^2/\Mpl^2}$. 
The tachyonic instability of the GR branch can drive the NS toward a scalarized final state endowed with both dark scalar and electric charges.
Observational constraints on such charges of NSs can be obtained from precise 
measurements of the orbital-period decay in binary pulsars.
In particular, precise timing observations of systems such as 
the Hulse-Taylor pulsar PSR B1913+16 provide stringent bounds on additional energy losses beyond those predicted by GR~\cite{Hulse:1974eb,Taylor:1982zz,Weisberg:2010zz}. 
The absence of significant deviations from the GR prediction implies tight constraints on dipole radiation sourced by dark scalar and electric charges, thereby restricting the allowed differences in the charge-to-mass ratios of the binary components.

Recently, several authors have constructed ECOs with regular centers in ESM theories, without introducing any matter fluid at the level of the Lagrangian \cite{Herdeiro:2019mbz,DeFelice:2024ops,DeFelice:2025vef,DeFelice:2026cse}.
If the coupling function $\mu(\phi)$ remains finite at all radii, there exist no asymptotically flat, static, and spherically symmetric ECO solutions~\cite{Herdeiro:2019yvb}.
This no-go theorem can be circumvented by allowing 
$\mu(\phi)$ to diverge at the center ($r=0$), while all physical quantities remain finite~\cite{Herdeiro:2019mbz,DeFelice:2024ops,DeFelice:2025vef,DeFelice:2026cse}.
A representative example is provided by the coupling
$\mu(\phi)=\mu_0+\mu_1 \Mpl^p/
(\phi-\phi_0)^p$ with $p>0$ \cite{DeFelice:2025vef}, where
$\phi_0$ denotes the scalar-field value at the center.
In such scenarios, electrically charged ECOs exhibit a shell-like structure, with the density peaking at an intermediate radius and vanishing both at $r \to 0$ and at spatial infinity.
Since these objects do not interact directly with SM particles, their existence must be probed through gravitational effects induced by their distinctive density profiles, such as photon rings, gravitational lensing, and the properties of innermost stable circular orbits \cite{DeFelice:2026cse}.
An alternative and complementary approach is to study gravitational waveforms emitted by binaries composed of ECOs.
This is precisely the focus of the present paper.

In this paper, we compute inspiral gravitational waveforms emitted by compact binary systems in ESM theories with a general 
coupling $\mu(\phi)$. 
The binaries may consist of BHs, NSs, or ECOs. 
We adopt an effective point-particle description, in which each compact object is characterized by its mass, electric charge, and scalar charge. 
GWs from electrically charged compact binaries have been investigated in several earlier works. 
In Einstein-Maxwell theories, charged BH binaries have been studied using analytical, post-Newtonian (PN), 
and numerical-relativity methods 
\cite{Cardoso:2016olt,Zhu:2018tzi,Liu:2020vsy,Liu:2020cds,Christiansen:2020pnv,Bozzola:2020mjx,Bozzola:2021elc,Benavides-Gallego:2022dpn}. 
These works showed how electric charges affect the conservative dynamics, radiation reaction, merger time, and inspiral-merger waveforms. 
Such waveform corrections have also been used to constrain BH charge-to-mass ratios with GW150914 and the GWTC-1 catalog \cite{Cardoso:2016olt,Wang:2020ori}. 
Higher-order PN corrections to the orbital evolution of charged BH binaries have also been derived 
\cite{Gupta:2022spq,Placidi:2025xyi}.
These studies mainly focus on electrically charged BHs in Einstein-Maxwell-type theories, or on hidden/dark $U(1)$ charges. 
To our knowledge, however, a systematic derivation of inspiral waveforms in ESM theories with a general coupling $\mu(\phi)$, including the effects of both electric and scalar charges for BHs, NSs, and ECOs, has not been presented.

Our calculation is organized by separating the near-zone dynamics from the far-zone radiation. 
The near-zone solutions determine the conservative binary dynamics and the effective gravitational coupling, whereas the far-zone solutions determine the tensor waveform and the energy flux carried by tensor, scalar, and vector radiation. 
Unlike in nonminimally coupled scalar-tensor theories, neither scalar nor vector modes couple directly to the tensor sector at linear order. As a result, they do not appear as additional components of the observed metric GW signal, but they modify the inspiral 
gravitational waveform through additional radiation reaction.
Using the stationary phase approximation, we obtain analytic expressions for the gravitational waveform in the frequency domain. 
We show that the leading correction to the waveform arises from dipole radiation sourced by differences in the electric and scalar charges of the two bodies. 
This effect can be encapsulated in a single parameter $b$, which governs both the amplitude and phase corrections to the waveform.

We also estimate the parameter $b$ for compact binaries composed of BH-BH, NS-NS, and ECO-ECO systems. 
For NS-NS binaries, the orbital-period decay of binary pulsars places stringent bounds on differences in the charge-to-mass ratios. We evaluate the orbital-period decay for a general coupling $\mu(\phi)$ in ESM theories and show that $b$ again serves as a key parameter characterizing the decay rate.
This enables us to constrain differences in the charge-to-mass ratios for NS-NS systems, whereas the same pulsar-timing bounds cannot be directly applied to BH-BH or ECO-ECO binaries. For the latter, one must instead rely on observational data from inspiral gravitational waveforms to extract information about dark charges.
Our general framework provides a useful approach to probing the existence of dark radiation and possible 
scalar-vector couplings in the vicinity of compact objects.

This paper is organized as follows. 
In Sec.~\ref{SecNear}, we introduce ESM theories and derive the near-zone solutions relevant to the conservative dynamics of compact binaries. 
In Sec.~\ref{SecTen}, we study GW propagation in the far, or wave, zone and derive the time-domain tensor waveform emitted by inspiraling binaries. 
In Sec.~\ref{Fresec}, we compute the energy flux carried by tensor, scalar, and vector radiation, evaluate the corresponding energy loss of the binary system, and derive the orbital-frequency evolution including the effects of dipole radiation. 
In Sec.~\ref{Fouriersec}, we obtain the frequency-domain waveform using the stationary phase approximation and discuss how the parameter $b$ affects the waveform amplitude and phase. 
In Sec.~\ref{Secapp}, we estimate the magnitude of $b$ for BH-BH, ECO-ECO, and NS-NS binaries that can arise in ESM theories. 
Finally, in Sec.~\ref{consec}, we summarize our findings and discuss their implications for GW observations and constraints on dark charges in ESM theories. 
Technical details of the scalar and vector radiation are presented in the Appendix. 
Throughout this paper, we work in natural units with $c=\hbar=1$.

\section{Near-zone dynamics in ESM theories}
\label{SecNear}

In this section, we set up the near-zone dynamics of compact binaries in ESM theories. 
This part of the analysis determines the conservative interaction between the two bodies 
and provides the orbital relation that will later enter both the time-domain waveform 
and the radiation-reaction equation.

The theories considered in this paper are described by the action
\begin{equation}
{\cal S}
=\int {\rm d}^4x\,\sqrt{-g}
\left[
\frac{\Mpl^2}{2}R + X + \mu(\phi)F
\right]
+{\cal S}_m,
\label{action}
\end{equation}
where $g$ is the determinant of the metric tensor 
$g_{\mu \nu}$, and
\begin{equation}
X = -\frac12 \nabla_\mu\phi \nabla^\mu\phi,
\qquad
F = -\frac14 F_{\mu\nu}F^{\mu\nu},
\qquad
F_{\mu\nu} = \nabla_\mu A_\nu-\nabla_\nu A_\mu\,.
\end{equation}
The scalar field $\phi$ is coupled to the $U(1)$ gauge field $A_{\mu}$ through the field-dependent gauge kinetic term $\mu(\phi)F$.

We model a binary system of compact objects as a pair of point particles with 
primary electric charges $q_I$ and scalar-field-dependent 
masses $m_I(\phi)$, where $I=A,B$ labels each particle. 
The matter action in Eq.~(\ref{action}) is then given by
\begin{equation}
{\cal S}_m
=-\sum_{I=A,B} \int m_I(\phi)\, {\rm d}\tau_I
-\sum_{I=A,B} q_I \int A_\mu \, {\rm d}x_I^\mu ,
\label{Sm}
\end{equation}
where ${\rm d}\tau_I = \sqrt{-g_{\mu\nu} {\rm d}x_I^\mu {\rm d}x_I^\nu}$ 
is the proper-time element of the $I$-th compact object. 
Here, ${\rm d}x_I^\mu$ denotes its infinitesimal worldline displacement, which can be expressed as ${\rm d}x_I^\mu = u_I^\mu {\rm d}\tau_I$, 
with $u_I^\mu$ being the four-velocity.

It is worth noting that the total action ${\cal S}$, including ${\cal S}_m$, preserves the $U(1)$ gauge invariance associated with $A_\mu$. 
Under the transformation $A_\mu \to A_\mu + \nabla_\mu \chi$, the field strength 
$F_{\mu\nu}$ is invariant, while the worldline 
gauge-field coupling in Eq.~(\ref{Sm}) transforms as
\be
\sum_{I} q_I \int A_\mu \, {\rm d}x_I^\mu 
\;\to\;
\sum_{I} q_I \int A_\mu \, {\rm d}x_I^\mu 
+ \sum_{I} q_I \int \nabla_\mu \chi \, {\rm d}x_I^\mu\,.
\ee
The second term is a total derivative along each worldline,
$\int \nabla_\mu \chi \, {\rm d}x_I^\mu 
= \chi \big|_{\rm endpoints}$
and hence contributes only a boundary term. 
For physically relevant configurations, this does not affect the dynamics, 
so that the action ${\cal S}$ remains $U(1)$ gauge invariant.

The scalar-field dependence of $m_I(\phi)$ in the matter action (\ref{Sm}) arises from secondary scalar hair induced by the coupling $\mu(\phi)F$. The presence of both primary electric charge and secondary scalar charge modifies the conservative dynamics of the binary and introduces additional radiation channels, including scalar and vector dipole emission. In the near zone, the fields are quasi-static and determine the binary potential, while in the far zone their radiative components govern the energy flux. As we show below, this separation establishes a direct link between the charge-dependent conservative dynamics and the phase evolution of the observed tensor waveform.

For a general spacetime background described 
by the line element
$\rd s^2 = g_{\mu\nu} \rd x^{\mu} \rd x^{\nu}$, 
with $x^{\mu}=(t,x^i)$, variation of the action~(\ref{action}) with respect to $g^{\mu\nu}$ yields
\begin{equation}
\Mpl^2 G_{\mu\nu}
= T_{\mu\nu}^{(\phi F)} + T_{\mu\nu}^{(m)}\,,
\label{EinsteinEq}
\end{equation}
where $G_{\mu\nu}$ is the Einstein tensor. 
The energy-momentum tensors associated with the 
scalar--gauge-field sector
and the matter sector are given by
\begin{align}
T_{\mu\nu}^{(\phi F)} &= 
\nabla_\mu \phi \, \nabla_\nu \phi
+ g_{\mu\nu} X
+ \mu(\phi) \left(
F_{\mu\alpha} F_{\nu}{}^{\alpha}
+ g_{\mu\nu} F
\right)\,, \nonumber \\
T_{\mu\nu}^{(m)} &= 
-\frac{2}{\sqrt{-g}}
\frac{\delta \mathcal{S}_m}{\delta g^{\mu\nu}}\,.
\label{TmunuDef}
\end{align}
More explicitly, the matter energy-momentum tensor 
obtained by varying the action~(\ref{Sm}) can be written as
\begin{equation}
T_{\mu\nu}^{(m)}(x)
= \sum_I
\frac{m_I(\phi)}{\sqrt{-g}}
\int \mathrm{d}\tau_I \,
u_{I\mu} u_{I\nu} \,
\delta^{(4)}\!\left(x - x_I \right),
\label{TmunuMatterPP}
\end{equation}
where $\delta^{(4)}(x-x_I)
=\delta(t-t_I)\,\delta^{(3)}(\bm{x}-\bm{x}_I)$ 
denotes the four-dimensional Dirac delta function.

Varying the action~(\ref{action}) with respect to $\phi$ yields
\begin{equation}
\Box\phi
+\mu_{,\phi}(\phi)F
=\frac{1}{\sqrt{-g}}
\sum_{I} \int {\rm d}\tau_I\,m_{I,\phi}(\phi)
\delta^{(4)}(x-x_I)\,,
\label{phieom1}
\end{equation}
where $\Box\phi \equiv \nabla_\mu \nabla^\mu \phi$, 
$\mu_{,\phi} \equiv {\rm d}\mu/{\rm d}\phi$, and
$m_{I,\phi} \equiv {\rm d}m_I/{\rm d}\phi$. 
Since the four-velocity of the $I$-th particle is 
$u_I^{\mu} = {\rm d}x_I^{\mu}/{\rm d}\tau_I$, 
its temporal component is given by 
$u_I^{0} = {\rm d}t/{\rm d}\tau_I$, so that 
${\rm d}\tau_I={\rm d}t/u_I^{0}$. 
One can then express Eq.~(\ref{phieom1}) as
\begin{equation}
\Box\phi+\mu_{,\phi}(\phi)F
=J_\phi\,,
\label{phieom2}
\end{equation}
where 
\begin{equation}
J_\phi \equiv \frac{1}{\sqrt{-g}}
\sum_{I} \frac{m_{I,\phi}(\phi)}{u_I^0}\,
\delta^{(3)}(\bm{x}-\bm{x}_I)\,.
\end{equation}
Varying the action~(\ref{action}) with respect to $A_\mu$ 
and using ${\rm d}x_I^\mu = u_I^\mu\,{\rm d}\tau_I$, we obtain
\begin{equation}
\nabla_\nu \left[ \mu(\phi)F^{\nu\mu} \right]
=\frac{1}{\sqrt{-g}} \sum_{I} q_I \int {\rm d}\tau_I\,u_I^\mu
\delta^{(4)}(x-x_I)\,.
\label{Aeom4d}
\end{equation}
This equation can be expressed as 
\be
\nabla_\nu \left[ \mu(\phi)F^{\nu\mu} \right]
=J_{A}^{\mu}\,,
\label{Fmunueq}
\ee
where 
\be
J_{A}^{\mu} \equiv 
\frac{1}{\sqrt{-g}}
\sum_{I} q_I \frac{u_I^\mu}{u_I^0}\,
\delta^{(3)}(\bm{x}-\bm{x}_I)\,.
\label{JAmu}
\ee
The currents $J_\phi$ and $J_A^\mu$ defined above are 
localized on the worldlines of the binary constituents and act as the sources for the scalar and vector fields, respectively. In the context of compact binary systems, it is useful to distinguish between two physically distinct regions: the near zone and the far (or wave) zone. 

The near zone is defined as the region surrounding the binary with a characteristic 
distance $r$ from the center of mass much smaller than the gravitational wavelength 
$\lambda_{\rm GW}$, i.e., $r \ll \lambda_{\rm GW}$. 
In this regime, the radiation time scale is much longer than the orbital time scale, 
so that the metric perturbation $h_{\mu\nu}$, as well as the scalar and vector fields 
$\phi$ and $A_\mu$, can be treated as quasi-static to leading order. 
As a result, retardation effects can be neglected, and the field equations reduce to elliptic-type equations sourced instantaneously by the currents $J_\phi$ and $J_A^\mu$. 
This allows one to derive the effective two-body 
interaction, including the conservative dynamics governing the orbital motion.

On the other hand, the far zone corresponds to distances $r \gg \lambda_{\rm GW}$, where the fields propagate as waves and carry energy away from the system. 
In this region, retardation effects are essential, and the solutions depend on the retarded time. 
The far-zone analysis is therefore relevant for computing the emitted radiation and constructing the gravitational waveforms observed at large distances.

In the rest of this section, we focus on the near-zone dynamics of the system and derive the quasi-static solutions for the metric perturbation, scalar, and vector fields. 
These solutions will be used to obtain the effective interaction potentials and the conservative energy of the binary system.

\subsection{Near-zone solutions}
\label{nearzonesec}

We work in the weak-field, slow-motion, and near-zone approximations around 
Minkowski spacetime with the metric 
$\eta_{\mu\nu} = \mathrm{diag}(-1,1,1,1)$, for which the line element is 
$\mathrm{d}s^2 = -\mathrm{d}t^2 + \delta_{ij}\,\mathrm{d}x^i \mathrm{d}x^j$.
We consider the metric perturbation $h_{\mu\nu}$ defined by
\begin{equation}
g_{\mu\nu}=\eta_{\mu\nu}+h_{\mu\nu}.
\label{gmunu}
\end{equation}
We expand the scalar field $\phi$ around its asymptotic value $\phi_\infty$ as
\begin{equation}
\phi=\phi_\infty+\varphi\,.
\end{equation}
In doing so, we assume that the background 
scalar-field value entering the point-particle description of BHs, NSs, and ECOs can be identified with
$\phi_\infty$, even in the near zone. 
This approximation is justified when the
internal scalar-field profiles of the compact objects are integrated out and
encoded in their scalar charges.
For the vector field $A_\mu$, we assume a configuration sourced by the electric charges of the binary. 
In the near zone, this gives rise to a Coulomb-type 
potential $A_0$, whereas the spatial components $A_i$ are velocity-suppressed and hence subleading.

We expand the scalar-dependent mass around $\phi_\infty$ as
\begin{equation}
 m_I(\phi)=m_I(\phi_\infty) \left[1+\alpha_I\left(\frac{\varphi}{M_{\rm Pl}}\right)
 +\frac12(\alpha_I^2+\beta_I)\left(\frac{\varphi}{M_{\rm Pl}}\right)^2+\cdots\right],
\label{massExpansion}
\end{equation}
where 
\begin{equation}
\alpha_I \equiv M_{\rm Pl} \left.\frac{\mathrm{d} \ln m_I}{\mathrm{d} \phi}\right|_{\phi=\phi_\infty},
\qquad
\beta_I \equiv M_{\rm Pl}^2\left.\frac{\mathrm{d}^2\ln m_I}
{\mathrm{d} \phi^2}\right|_{\phi=\phi_\infty}.
\label{alphaBetaDef}
\end{equation}
Keeping only the linear term in $\varphi$ is sufficient for the Newtonian-order conservative dynamics in the present analysis. 
We therefore substitute
$m_I(\phi) = m_I(\phi_\infty) \left(1 + \alpha_I \varphi / M_{\rm Pl}\right)$ into the matter action~(\ref{Sm}). 
In the following, we denote $m_I(\phi_\infty)$ simply by $m_I$ for brevity.

Expanding the proper time 
${\rm d}\tau_I = \sqrt{-g_{\mu\nu} \, {\rm d}x_I^\mu {\rm d}x_I^\nu}$ 
around the Minkowski background, we obtain
\begin{equation}
{\rm d}\tau_I = {\rm d}t \sqrt{1-v_I^2-h_{00}+\mathcal{O}(\epsilon^2)},
\end{equation}
where $v_I^2 \equiv \delta_{ij} v_I^i v_I^j$ with
$v_I^i \equiv {\rm d}x_I^i / {\rm d}t$, and $\epsilon$ 
denotes the order of perturbations. 
Here, terms involving $h_{0i}$ and $h_{ij}$ have been neglected because they enter only at higher PN orders beyond the Newtonian approximation considered here.
In the weak-field and slow-motion regime, $v_I^2 \ll 1$ and $|h_{\mu\nu}| \ll 1$, this expression can be further approximated as
${\rm d}\tau_I = {\rm d}t \left[1 - v_I^2/2 - h_{00}/2 
+ \mathcal{O}(\epsilon^2) \right]$.
Retaining only the leading couplings relevant at Newtonian order 
and discarding irrelevant constant terms, the matter action reduces to
\begin{equation}
 {\cal S}_m \simeq \int {\rm d}t\, \sum_I \left[
 \frac12 m_I v_I^2 -V({\bm x}_I) \right]\,,
\label{SmNR}
\end{equation}
where 
\begin{equation}
V({\bm x}_I)=-\frac12 m_I h_{00}(\bm{x}_I) 
+\alpha_I m_I  \frac{\varphi(\bm{x}_I)}{M_{\rm Pl}} 
+q_I A_0(\bm{x}_I)\,.
\label{potential}
\end{equation}
Equation~(\ref{SmNR}) corresponds to the standard nonrelativistic action 
for a system of point particles. The first term, $m_I v_I^2/2$, 
represents the kinetic energy of the $I$-th body in the binary system, 
while $V({\bm x}_I)$ plays the role of an effective potential 
that encodes the interactions mediated by the metric, scalar, 
and vector fields. In particular, the term proportional to $h_{00}$ 
should reproduce the Newtonian gravitational potential, whereas the 
scalar and vector contributions give rise to additional long-range forces.

Let us first derive $h_{00}({\bm x}_I)$
in the weak-field and static limits, starting from the linearized Einstein equations. 
We focus on the leading-order contribution sourced 
by the rest-mass density in the near zone. 
Then, the leading contribution to the metric perturbation is sourced by the matter energy-momentum tensor $T_{\mu\nu}^{(m)}$. 
The scalar and vector contributions are treated separately as additional 
interaction potentials. Thus, at leading order, we approximate
\begin{equation}
G_{\mu\nu} \simeq 8\pi G_N T_{\mu\nu}^{(m)}\,,
\end{equation}
where $G_N = (8\pi M_{\rm Pl}^2)^{-1}$ is the Newton gravitational constant.  
At first linear order in $h_{\mu\nu}$, the Einstein tensor becomes
\begin{equation}
G_{\mu\nu}^{(1)} = -\frac{1}{2}\Box \bar h_{\mu\nu}
-\frac{1}{2}\eta_{\mu\nu}\partial^\alpha\partial^\beta \bar h_{\alpha\beta}
+\partial^\alpha\partial_{(\mu}\bar h_{\nu)\alpha}\,,
\end{equation}
where 
\begin{equation}
\bar h_{\mu\nu}=h_{\mu\nu}-\frac{1}{2}\eta_{\mu\nu}h\,,
\label{barh}
\end{equation}
with the trace $h \equiv \eta^{\alpha \beta}h_{\alpha \beta}$. 
Imposing the harmonic gauge condition $\partial^\mu \bar h_{\mu\nu}=0$, 
this simplifies to $G_{\mu\nu}^{(1)}=-\Box \bar h_{\mu\nu}/2$.
Therefore, the Einstein equation reduces to
\begin{equation}
 \Box \bar h_{\mu\nu}
 =-16\pi G_N T_{\mu\nu}^{(m)}\,.
 \label{linhappA}
\end{equation}
In the weak-field, nonrelativistic limit, the dominant component to 
$T_{\mu \nu}^{(m)}$ in Eq.~(\ref{TmunuMatterPP}) is
\begin{equation}
T_{00}^{(m)} \simeq \rho_m({\bm x})\,,
\end{equation}
where $\rho_m(\bm{x})$ is the matter density defined by 
\begin{equation}
\rho_m(\bm{x})=\sum_I m_I\,\delta^{(3)}(\bm{x}-\bm{x}_I)\,.
\end{equation}
Hence, the $(\mu,\nu)=(0,0)$ component of 
Eq.~\eqref{linhappA} gives
$\Box \bar h_{00}=-16\pi G_N \rho_m$.
In the near zone, time derivatives are subleading compared to spatial derivatives, 
so that $\Box$ can be replaced with $\nabla^2$. 
It then follows that 
\begin{equation}
\nabla^2 \bar h_{00}(\bm{x})=
-16\pi G_N \rho_m(\bm{x})\,.
\label{h00eq}
\end{equation}
The solution to this equation is given by 
\begin{equation}
\bar{h}_{00}(\bm{x})
=4G_N \int \rd^3x' \frac{\rho_m (\bm{x}')}
{|\bm{x}-\bm{x}'|}
=4G_N \sum_I \frac{m_I}{|\bm{x}-\bm{x}_I|}\,.
\label{h0so}
\end{equation}
At the position of the $I$-th compact object, 
the metric perturbation $h_{00}(\bm{x}_I)$ is given by
\begin{equation}
h_{00}(\bm{x}_I)
=2G_N \sum_{J \neq I} \frac{m_J}{|\bm{x}_I-\bm{x}_J|}
=\frac{2G_N m_J}{r}\,,
\label{h0so1}
\end{equation}
where $r=|\bm{x}_I-\bm{x}_J|$, and the second equality holds for a binary system. 
Substituting this expression into the first term of 
Eq.~(\ref{potential}), we obtain
\begin{equation}
V_g({\bm{x}}_I)
\equiv -\frac12 m_I h_{00}({\bm{x}}_I)
= -\frac{G_N m_I m_J}{r}\,,
\label{Vg}
\end{equation}
which corresponds to the Newtonian gravitational potential.

To compute $\varphi (\bm{x}_I)$ and  $A_0 (\bm{x}_I)$ in Eq.~(\ref{potential}), 
we express these fields by using the three-dimensional delta function as 
\begin{equation}
\varphi(\bm{x}_I)=\int \mathrm{d}^3x\,\varphi(\bm{x})\,\delta^{(3)}(\bm{x}-\bm{x}_I),
\qquad
A_0(\bm{x}_I)=\int \mathrm{d}^3x\,A_0(\bm{x})\,\delta^{(3)}(\bm{x}-\bm{x}_I).
\end{equation}
Then, the scalar and vector couplings in the action (\ref{SmNR}) 
can be written in the standard field-theory form:
\begin{equation}
 {\cal S}_{m,{\rm source}} \simeq -\int {\rm d}t\, {\rm d}^3 x
 \left[ \rho_s(\bm{x}) \frac{\varphi(\bm{x})}{M_{\rm Pl}}
 +\rho_e(\bm{x})\,A_0(\bm{x})
 \right],
 \label{Sms}
\end{equation}
where 
\begin{equation}
\rho_s(\bm{x}) \equiv \sum_I \alpha_I m_I \delta^{(3)}(\bm{x}-\bm{x}_I),
\qquad
\rho_e(\bm{x}) \equiv \sum_I q_I\,\delta^{(3)}(\bm{x}-\bm{x}_I).
\label{rhos}
\end{equation}
Here, $\rho_s(\bm{x})$ and $\rho_e(\bm{x})$ correspond to 
the scalar and electric source densities, respectively.

For the scalar field, the scalar-kinetic action $\sqrt{-g}\,X$ 
in the action (\ref{action}) also 
contributes to the equation of motion for $\varphi$. 
Under the near-zone approximation, the corresponding action for 
a static scalar field $\varphi({\bm x})$ takes the form 
\be
{\cal S}_{\varphi}=\int \rd t\,\rd^3x
\left[-\frac{1}{2}(\nabla\varphi)^2
-\rho_s\frac{\varphi}{\Mpl} \right]\,.
\label{Sphi}
\ee
One may wonder whether the coupling $\mu(\phi)F$ also contributes at this order.
Expanding $\mu(\phi)$ around $\phi=\phi_{\infty}$ in the presence of a
background Coulomb field indeed generates a term 
linear in $\varphi$.
In the point-particle description used here, however, this effect is already encoded in the $\phi$-dependence of the masses $m_I(\phi)$, or equivalently in the scalar source density $\rho_s$. We therefore do not include the contribution arising from $\mu_{,\phi}\varphi F$ as an independent source in 
the near-zone scalar equation. In any case, this contribution 
is subdominant relative to the leading point-particle 
scalar source.

Varying the action~(\ref{Sphi}) with respect to $\varphi$, we obtain
\begin{equation}
 \nabla^2 \varphi({\bm x})
 = \frac{\rho_s({\bm x})}{M_{\rm Pl}}
 = \frac{1}{M_{\rm Pl}} \sum_I \alpha_I m_I \,
 \delta^{(3)}({\bm x} - {\bm x}_I)\,,
 \label{Poissonphi}
\end{equation}
which can also be recovered from Eq.~(\ref{phieom2}) in the static, 
weak-field limit around a Minkowski background. 
The solution to Eq.~(\ref{Poissonphi}) is
\begin{equation}
\varphi({\bm x})
= -\frac{1}{4\pi M_{\rm Pl}}
\int \mathrm{d}^3 x' \,
\frac{\rho_s({\bm x}')}{|{\bm x} - {\bm x}'|}
= -\frac{1}{4\pi M_{\rm Pl}}
\sum_I \frac{\alpha_I m_I}{|{\bm x} - {\bm x}_I|}\,.
\label{phiSolDetailed}
\end{equation}
Evaluating this expression at the position ${\bm x}_I$, we find
\begin{equation}
\varphi({\bm x}_I)
= -\frac{1}{4\pi M_{\rm Pl}}
\sum_{J \neq I}
\frac{\alpha_J m_J}{|{\bm x}_I - {\bm x}_J|}
= -\frac{1}{4\pi M_{\rm Pl}} \frac{\alpha_J m_J}{r}\,.
\label{varphixI}
\end{equation}
Substituting this into the second term of Eq.~(\ref{potential}), 
the scalar-mediated potential is given by
\begin{equation}
V_{\varphi}({\bm x}_I)
\equiv
\alpha_I m_I \frac{\varphi({\bm x}_I)}{M_{\rm Pl}}
= -\frac{2 G_N \alpha_I \alpha_J m_I m_J}{r}\,,
\label{Vvarphi}
\end{equation}
where we have used the relation $M_{\rm Pl}^{-2} = 8\pi G_N$. 
As long as both $\alpha_I$ and $\alpha_J$ are nonvanishing, 
the scalar field mediates an additional long-range force 
on top of the Newtonian gravitational interaction.

In the vector sector, under the static approximation, the only nonvanishing
components of $F_{\mu\nu}$ are $F_{0i}=-F_{i0}=-\partial_i A_0$, while the
magnetic components vanish. This gives $F=(\nabla A_0)^2/2$. As in the scalar
sector, we neglect higher-order corrections arising from the expansion of
$\mu(\phi)$ around $\phi_\infty$ and retain 
only its leading value
$\mu_\infty\equiv \mu(\phi_\infty)$ in the near-zone 
vector action.
The vector action for a purely electric, static configuration is then given by
\begin{equation}
{\cal S}_A = \int {\rm d}t\, {\rm d}^3x
\left[\frac{\mu_\infty}{2}(\nabla A_0)^2 - \rho_e A_0\right].
\label{SAStart}
\end{equation}
Varying this action with respect to $A_0$, we obtain
\begin{equation}
 \nabla^2 A_0 = -\frac{1}{\mu_\infty}\rho_e
 = -\frac{1}{\mu_\infty}\sum_I q_I\,\delta^{(3)}(\bm{x}-\bm{x}_I)\,.
\label{PoissonA0}
\end{equation}
Equation~\eqref{PoissonA0} is solved by
\be
 A_0(\bm{x}) = \frac{1}{4\pi\mu_\infty}\int {\rm d}^3x'\,\frac{\rho_e(\bm{x}')}{|\bm{x}-\bm{x}'|}
 = \frac{1}{4\pi\mu_\infty}\sum_I \frac{q_I}{|\bm{x}-\bm{x}_I|}\,.
 \label{A0SolDetailed}
\ee
This solution corresponds to the usual electrostatic potential sourced by the charges, 
with $\mu_\infty$ determining the effective coupling strength. 

At the position $\bm{x}_I$, Eq.~(\ref{A0SolDetailed}) reduces to 
\be
 A_0(\bm{x}_I) = \frac{1}{4\pi\mu_\infty}
 \sum_{J \neq I} \frac{q_J}{|\bm{x}_I-\bm{x}_J|}
 =\frac{1}{4\pi\mu_\infty} \frac{q_J}{r}\,.
\ee
Therefore, the vector-mediated potential in Eq.~(\ref{potential}) yields 
\be
V_{A_0}=q_I A_0(\bm{x}_I)=\frac{1}{4\pi\mu_\infty} 
\frac{q_I q_J}{r}\,.
\label{VA0}
\ee
From Eqs.~(\ref{Vg}), (\ref{Vvarphi}), and (\ref{VA0}), the total potential 
(\ref{potential}) is expressed as 
\begin{equation}
 V(r)= -\frac{G_N m_I m_J}{r} \left( 1+2\alpha_I \alpha_J 
 \right)+\frac{1}{4\pi\mu_\infty} \frac{q_I q_J}{r}\,.
\label{Vsum1}
\end{equation}
We define the dimensionless vector charge-to-mass ratio
\begin{equation}
 \sigma_I \equiv \frac{q_I}{\sqrt{4\pi\mu_\infty G_N}\,m_I}\,.
\label{sigmaDefApp}
\end{equation}
Then, Eq.~(\ref{Vsum1}) can be written as 
\begin{equation}
V(r)= -\frac{G_{\rm eff} m_I m_J}{r}\,,
\label{Vsum2}
\end{equation}
where 
\be
G_{\rm eff}=G_N\left(1+2\alpha_I\alpha_J-\sigma_I\sigma_J \right)\,.
\ee
This shows that the conservative binary dynamics is governed by the effective Newtonian coupling $G_{\rm eff}$.
The scalar-mediated interaction is attractive for $\alpha_I\alpha_J>0$, whereas the vector-mediated interaction is repulsive for like charges and attractive for opposite charges.

\subsection{Circular orbits}

In Sec.~\ref{nearzonesec}, we showed that the near-zone binary dynamics is described by the matter action~(\ref{SmNR}) with the effective potential
given in Eq.~(\ref{Vsum2}). 
For a binary system composed of two compact
objects labeled by $A$ and $B$, the matter action can be written as
${\cal S}_m=\int {\rm d}t\,{\cal L}_{\rm eff}$, where
\be
{\cal L}_{\rm eff}=\frac{1}{2}m_A v_A^2+\frac{1}{2}m_B v_B^2
+\frac{G_{\rm eff}\,m_A m_B}{r}\,.
\label{Leff}
\ee
We now consider the conservative dynamics of compact binaries in circular orbits.
We introduce the standard mass parameters
\begin{equation}
M \equiv m_A+m_B\,,
\qquad
m_{r} \equiv \frac{m_A m_B}{M}\,,
\qquad
\eta \equiv \frac{m_{r}}{M}\,,
\qquad
M_c \equiv \eta^{3/5}M\,,
\label{eta}
\end{equation}
where $m_{r}$ is the reduced mass. 
Introducing the center-of-mass coordinate 
${\bm R}=(m_A\bm{x}_A+m_B\bm{x}_B)/M$ 
and the relative coordinate ${\bm r}={\bm x}_A-{\bm x}_B$, 
the kinetic term becomes 
$K=M \dot{\bm R}^2/2+m_{r} \dot{\bm r}^2/2$. 
In the center-of-mass frame ($\dot{\bm R}={\bm 0}$), this reduces to 
$K=m_{r} \dot{\bm r}^2/2$. 

For circular orbits with constant separation $r=|{\bm r}|$, 
we have $\dot{\bm r}^2=r^2\omega^2$, so that 
$K=m_{r} r^2\omega^2/2$, where $\omega$ is the orbital angular frequency. 
The effective Lagrangian (\ref{Leff}) 
then becomes
\begin{equation}
{\cal L}_{\rm eff}=\frac12 m_{r} r^2\omega^2
+\frac{G_{\rm eff} M m_{r}}{r}\,.
\label{Lrel}
\end{equation}
Varying this Lagrangian with respect to $r$, we obtain
\begin{equation}
r \omega^2=\frac{G_{\rm eff} M}{r^2}\,.
\end{equation}
The orbital velocity is therefore
\begin{equation}
v=r\omega=\sqrt{\frac{G_{\rm eff} M}{r}}
=(G_{\rm eff} M\omega)^{1/3}\,.
\label{vrel1}
\end{equation}
The orbital energy of the binary system is
\be
E=\frac12 m_{r} v^2-\frac{G_{\rm eff} M m_{r}}{r}
=-\frac12 m_{r} v^2 
=-\frac12 m_{r}(G_{\rm eff} M\omega)^{2/3}\,.
\label{Eene}
\ee
The assumption of circular orbits is well justified for compact binaries emitting 
radiation. Gravitational (and scalar/vector) radiation reaction enters at higher PN order and drives a gradual inspiral, while efficiently damping 
the orbital eccentricity. As a result, by the time the binary enters the 
frequency band relevant for GW observations, the orbit is expected 
to be very close to circular.

In the conservative approximation, where radiation reaction is neglected, 
$\omega$ can be treated as constant in time, and the energy $E$ is conserved. 
In Sec.~\ref{SecTen}, we derive the gravitational waveform in the time domain 
under this approximation. In Sec.~\ref{Fresec}, we include the effect of 
radiation reaction and show that $\omega$ increases in time due to energy loss.


The near-zone analysis provides the conservative dynamics and orbital relations 
that determine the motion of the binary system. 
To compute the gravitational radiation emitted by the system, however, 
one must solve the field equations in the far (wave) zone, where the fields propagate as outgoing waves. 
In the next section, we derive the time-domain waveform observed at large distances 
by performing this far-zone analysis.

\section{Tensor waveform in the time domain}
\label{SecTen}

In this section, we investigate the propagation of GWs in the far (wave) zone 
and derive the time-domain waveform emitted by inspiraling compact binaries 
in ESM theories. 
As discussed in Sec.~\ref{SecNear}, the near-zone analysis determines the conservative dynamics 
and orbital motion of the binary, which provide the source of radiation. 
In contrast to the near zone, where the fields can be treated as quasi-static, 
the far zone corresponds to distances much larger than the GW wavelength, 
where the radiative degrees of freedom propagate as outgoing waves. 
By solving the field equations with retarded Green's functions, we obtain the tensor radiative field observed at a distance $D$ from the source, 
with all source quantities evaluated at the retarded time.

Since the scalar and vector perturbations do not couple to the transverse-traceless 
tensor modes at linear order in the far zone, the corresponding wave equations 
can be solved independently. 
For completeness, we present the explicit forms of the scalar and vector wave solutions in Appendix. 
The tensor result derived in this section will be used later to construct 
the frequency-domain waveform.

Let us consider the propagation of GWs on a Minkowski background, 
where the metric $g_{\mu\nu}$ is decomposed as in Eq.~(\ref{gmunu}). 
Introducing the trace-reversed field $\bar{h}_{\mu\nu}$ as in Eq.~(\ref{barh}), 
and choosing the harmonic gauge condition $\partial^\mu \bar{h}_{\mu\nu}=0$, 
the linearized Einstein equations reduce to Eq.~(\ref{linhappA}). 
The solution to Eq.~(\ref{linhappA}) at the observation point
$x^\mu=(t,\vb*{X})$ can be written as
\begin{equation}
\bar{h}_{\mu\nu}(x)
=16\pi G_N\int {\rm d}^4x'\,G_{\rm ret}(x-x')T_{\mu\nu}(x')\,,
\label{greenconvdetail}
\end{equation}
where $x'^\mu=(t',\vb*{x}')$ denotes the source point, and
\begin{equation}
G_{\rm ret}(t-t',\vb*{X}-\vb*{x}')
=\frac{\delta\!\left(t-t'-|\vb*{X}-\vb*{x}'|\right)}
{4\pi |\vb*{X}-\vb*{x}'|}
\,\Theta(t-t')\,,
\label{Gretflatdetail}
\end{equation}
describes propagation along the past light cone at unit speed and
vanishes for $t<t'$.

Substituting Eq.~\eqref{Gretflatdetail} into 
Eq.~\eqref{greenconvdetail}, we obtain
\begin{align}
\bar{h}_{\mu\nu}(t,\vb*{X})
&= 
4G_N\int {\rm d}t'\,{\rm d}^3 x'\,
\frac{\delta\!\left(t-t'-|\vb*{X}-\vb*{x}'|\right)}{|\vb*{X}-\vb*{x}'|}
T_{\mu\nu}(t',\vb*{x}') \nonumber\\
&= 4G_N\int {\rm d}^3 x'\,
\frac{T_{\mu\nu}(t-|\vb*{X}-\vb*{x}'|,\vb*{x}')}{|\vb*{X}-\vb*{x}'|}\,,
\label{retardedhmunu}
\end{align}
where, in the second line, we have performed the integration over $t'$ 
using the delta function, which fixes the source time to the retarded time 
$t'=t-|\vb*{X}-\vb*{x}'|$.

To obtain the leading radiative field, we evaluate this expression in the wave zone, where the observer is located far from the source. 
Introducing
\begin{equation}
D\equiv |\vb*{X}|,
\qquad
\hat{\vb*{n}}\equiv \frac{\vb*{X}}{D},
\end{equation}
and using $D\gg |\vb*{x}'|$, we expand the distance 
between the source and observer points as
\begin{equation}
|\vb*{X}-\vb*{x}'|=
D-\hat{\vb*{n}}\cdot\vb*{x}'+{\cal O}\left( \frac{|\vb*{x}'|^2}{D} \right)\,.
\label{bXx}
\end{equation}
For the leading quadrupole radiation, 
it is sufficient to keep only 
the lowest term in the amplitude factor, 
$|\vb*{X}-\vb*{x}'|\simeq D$, 
and to evaluate the source at the common 
retarded time $t-D$ at leading order. 
Then, the spatial components of the retarded 
solution reduce to
\begin{equation}
\bar h_{ij}(t,\vb*{X})\simeq \frac{4G_N}{D}
\int {\rm d}^3x'\,T_{ij}(t-D,\vb*{x}')\,.
\label{farzonehijdetail}
\end{equation}
This shows that the amplitude decreases as $1/D$, and all source points 
contribute at the same leading retarded time $t-D$.

We now relate the integral of $T_{ij}$ in Eq.~\eqref{farzonehijdetail} 
to the mass quadrupole moment defined by
\begin{equation}
I_{ij}(t)\equiv \int {\rm d}^3x'\,T^{00}(t,\vb*{x}')x'_i x'_j\,.
\label{Iij}
\end{equation}
Using energy-momentum conservation,
$\partial_\mu T^{\mu\nu}=0$, and 
assuming that the source is localized
so that surface terms vanish, we find, after taking two time derivatives
and integrating by parts,
\begin{equation}
\ddot I_{ij}
=2\int {\rm d}^3x'\,T^{ij}\,.
\label{Iijddotdetail}
\end{equation}
In the weak-field limit, the spatial indices can 
be raised and lowered with $\delta_{ij}$, so that $T^{ij}=T_{ij}$. 
Substituting Eq.~\eqref{Iijddotdetail} into Eq.~\eqref{farzonehijdetail}, we obtain
\begin{equation}
\bar h_{ij}(t,\vb*{X})\simeq 
\frac{2G_N}{D}\,\ddot I_{ij}(t-D)\,.
\label{hij}
\end{equation}
Since the tensor perturbations are governed by the Einstein sector, the leading-order radiative field takes the same form as in GR, with the important difference that the orbital dynamics entering the quadrupole moment is controlled by the effective gravitational 
coupling $G_{\rm eff}$ rather than $G_N$.

Let us consider a circular orbit of compact objects 
$A$ and $B$ in the $(x,y)$ plane, with a constant 
separation $r=|\vb*{x}_A-\vb*{x}_B|$ and a constant 
angular frequency $\omega$, such that
\begin{equation}
\vb*{r}(t_r)=\vb*{x}_A-\vb*{x}_B
= \left[ r \cos \Phi(t_r),r \sin \Phi(t_r),0 \right]\,,
\label{vbr}
\end{equation}
where $\Phi(t_r)=\omega t_r$ and 
$t_r=t-D$ is the retarded time. 
In the center-of-mass frame, the individual positions are 
$\vb*{x}_A=m_B \vb*{r}/M$ and $\vb*{x}_B=-m_A\vb*{r}/M$, 
with $M=m_A+m_B$. Then, the mass quadrupole moment 
of the binary system can be expressed as 
\begin{equation}
I_{ij}=m_A (x_A)_i (x_A)_j+m_B (x_B)_i (x_B)_j
=m_{r} r_i r_j\,,
\label{Iijpp}
\end{equation}
where $m_{r}$ is the reduced mass defined 
in Eq.~(\ref{eta}), and $r_i$ and $r_j$ 
are components of $\vb*{r}$. 
Therefore, when computing the leading quadrupole radiation, 
the binary system can be treated as a single effective 
particle of reduced mass $m_r$ moving on the 
relative orbit $\vb*{r}(t_r)$.
For the circular orbit above, the nonvanishing 
quadrupole components are
\be
I_{xx}=\frac12 m_{r} r^2\left(1+\cos2\Phi\right)\,,\qquad
I_{yy}=\frac12 m_{r} r^2\left(1-\cos2\Phi\right)\,,\qquad
I_{xy}=I_{yx}=\frac12 m_{r} r^2\sin2\Phi\,.
\label{Ixyz}
\ee
Hence, the radiative quadrupole varies at frequency $2\omega$, as expected 
for tensor waves from a circular binary. 
From Eq.~(\ref{hij}), the nonvanishing components of $\bar{h}_{ij}$ are given by 
\be
\bar{h}_{xx}=-\frac{4G_N}{D}m_{r} r^2 \omega^2 \cos(2\Phi)\,,\qquad 
\bar{h}_{yy}=\frac{4G_N}{D}m_{r} r^2 \omega^2 \cos(2\Phi)\,,\qquad 
\bar{h}_{xy}=\bar{h}_{yx}=-\frac{4G_N}{D}m_{r} r^2 \omega^2 
\sin(2\Phi)\,.
\label{barhcom}
\ee

Let us derive the transverse-traceless (TT) component 
of GWs at the observer point $\vb*{X}$. Using the axial symmetry of the circular orbit, we choose the azimuthal angle of the observer to be zero, so that the unit vector 
$\hat{\vb*{n}}=\vb*{X}/D$ pointing from the source 
to the observer can be written as
\be
\hat{\vb*{n}}=(\sin\iota,0,\cos\iota)\,,
\label{hatndef}
\ee
where $\iota$ is the inclination angle between 
the orbital angular momentum and the line of sight. 
It is convenient to introduce 
observer-frame coordinates $(X,Y,Z)$ 
such that the $Z$-axis is aligned 
with $\hat{\vb*{n}}$.
This is achieved by the rotation
\begin{equation}
X=x\cos\iota-z\sin\iota,\qquad
Y=y,\qquad
Z=x\sin\iota+z\cos\iota\,.
\end{equation}
Since the nonvanishing components of $\bar h_{ij}$ are 
given by Eq.~(\ref{barhcom}), the relevant components
in the observer frame are
\begin{align}
\bar h_{XX}
&=\cos^2\iota\,\bar h_{xx}
= -\frac{4G_N}{D}m_{r} r^2\omega^2
\cos^2\iota\,\cos(2\Phi),\\
\bar h_{YY}
&=\bar h_{yy}
= \frac{4G_N}{D}m_{r} r^2\omega^2
\cos(2\Phi),\\
\bar h_{XY}
&=\cos\iota\,\bar h_{xy}
= -\frac{4G_N}{D}m_{r} r^2\omega^2
\cos\iota\,\sin(2\Phi).
\end{align}
The TT projection of $\bar{h}_{ij}$ is given by
\begin{equation}
\bar h_{ij}^{\rm TT}
=
\Lambda_{ij,kl} \bar h_{kl}\,,
\end{equation}
where $\Lambda_{ij,kl}$ is the TT projector defined by 
\begin{equation}
\Lambda_{ij,kl}
=
P_{ik}P_{jl}
-\frac12 P_{ij}P_{kl},
\qquad
P_{ij}=\delta_{ij}-\hat n_i \hat n_j \,.
\label{Lambda}
\end{equation}
In the TT sector, the trace-reversed perturbation 
coincides with the metric perturbation itself, $\bar{h}_{ij}^{\rm TT}=h_{ij}^{\rm TT}$.
In the observer frame, where $\hat{\vb*{n}}$ 
is aligned with the $Z$-axis, the independent 
TT components are
\begin{align}
h_{XX}^{\rm TT}
&=
\frac12\left(\bar h_{XX}-\bar h_{YY}\right)
=
-\frac{4G_N}{D}m_{r} r^2\omega^2
\frac{1+\cos^2\iota}{2}\cos(2\Phi),\\
h_{XY}^{\rm TT}
&=
\bar h_{XY}
=
-\frac{4G_N}{D}m_{r} r^2\omega^2
\cos\iota\,\sin(2\Phi).
\end{align}
With the standard decomposition of the polarized modes of the TT waveform,
$h_+ = h_{XX}^{\rm TT}$ and $h_\times = h_{XY}^{\rm TT}$,
we obtain
\begin{align}
h_+(t)
&=
-\frac{4G_N}{D}
M_c^{5/3}(G_{\rm eff} \omega)^{2/3}
\frac{1+\cos^2\iota}{2}\cos(2\Phi)\,,
\label{hp} \\
h_\times(t)
&=
-\frac{4G_N}{D}
M_c^{5/3}(G_{\rm eff}\omega)^{2/3}
\cos\iota\,\sin(2\Phi)\,,
\label{hm} 
\end{align}
where we have used the relations $r=(G_{\rm eff} M/\omega^2)^{1/3}$, 
$m_{r}=\eta M$, and $M_c=\eta^{3/5}M$.
This expression represents the ESM counterpart of the time-domain 
tensor waveform. The effective gravitational coupling $G_{\rm eff}$ 
enters through its modification of the orbital separation $r$.

\section{Energy loss during inspiral}
\label{Fresec}

In this section, we evaluate the energy loss during 
the inspiral phase by computing
the radiated power carried by the tensor, scalar, and vector sectors.
This step provides the link between the far-zone radiation fields and the slow evolution of the near-zone orbit through the balance equation.
The radiated power associated with each sector is derived in
Subsecs.~\ref{Tenfsec},~\ref{Scafsec}, and~\ref{Vecfsec}, respectively.
In Subsec.~\ref{Totalfsec}, we combine these 
contributions to obtain the total radiated flux 
and derive the associated frequency sweep.
In Subsec.~\ref{orbital}, we discuss the resulting 
orbital-period decay of the binary system driven by radiation reaction.

Throughout this section, we consider a circular binary orbit with a fixed separation $r=|\vb*{x}_A-\vb*{x}_B|$ 
at leading order. The separation vector
$\vb*{r}(t_r)=\vb*{x}_A-\vb*{x}_B$, evaluated 
at the retarded time $t_r=t-D$, is given by Eq.~(\ref{vbr}).
The orbital angular frequency $\omega$ is also treated as constant at leading order, while its slow evolution due to radiation reaction is consistently
taken into account in Subsecs.~\ref{Totalfsec} and~\ref{orbital}.

\subsection{Tensor quadrupole flux}
\label{Tenfsec}

Although the orbital energy is determined by the near-zone dynamics of the
binary, its loss is most conveniently evaluated from the energy flux carried
by radiation in the far (wave) zone. We therefore use the far-zone tensor
waveform derived in Sec.~\ref{SecTen}. For a slowly moving binary, the TT
metric perturbation evaluated at the observer position is
\begin{equation}
h_{ij}^{\rm TT}(t,\vb*{X})
=\frac{2G_N}{D}\,\Lambda_{ij,kl}(\hat{\vb*{n}})\,\ddot I_{kl}(t-D),
\label{hTTfarzonefluxdetail}
\end{equation}
where $D=|\vb*{X}|$ is the distance to the observer,
$\hat{\vb*{n}}=\vb*{X}/D$ is the propagation direction, and
$\Lambda_{ij,kl}$ is the TT projector defined in Eq.~(\ref{Lambda}).

The energy carried by gravitational radiation in the 
far zone is described by the Isaacson effective 
stress tensor \cite{Isaacson:1968zzb}, which represents
the effective energy-momentum of high-frequency 
metric perturbations on a slowly varying background spacetime. For the TT tensor perturbation, 
it is given by
\begin{equation}
T_{\mu\nu}^{\rm GW}
=\frac{1}{32\pi G_N}
\left\langle \partial_\mu 
h_{ij}^{\rm TT}\partial_\nu h_{ij}^{\rm TT}
\right\rangle\,.
\label{IsaacsonstressfluxdetailTT}
\end{equation}
Here, the angle brackets denote an average over several wavelengths or periods of the GWs, performed over a spacetime region large compared to the wavelength 
but small compared to the characteristic curvature scale
of the background. This averaging procedure removes 
rapidly oscillating terms and isolates the effective 
energy-momentum of the radiative degrees
of freedom.

The total tensor power is obtained by integrating the outward radial energy flux over a sphere of radius 
$D$ in the far zone. With our metric signature,
the outward flux is given by $-T_{0D}^{\rm GW}$, 
so that
\begin{equation}
\mathcal{F}_{\rm T}
= -\int D^2 \, T_{0D}^{\rm GW} \, {\rm d}\Omega\,,
\label{FluxfromT0rfluxdetail}
\end{equation}
where ${\rm d}\Omega$ denotes the element of solid angle. Since the waveform
depends on the retarded time $t-D$, it follows that
$\partial_D h_{ij}^{\rm TT}=-\dot h_{ij}^{\rm TT}$ at leading order in the
far zone. Hence,
\begin{equation}
T_{0D}^{\rm GW}
=
\frac{1}{32\pi G_N}
\left\langle
\partial_0 h_{ij}^{\rm TT} \, \partial_D h_{ij}^{\rm TT}
\right\rangle
=
-\frac{1}{32\pi G_N}
\left\langle
\dot h_{ij}^{\rm TT} \dot h_{ij}^{\rm TT}
\right\rangle
=
-\frac{G_N}{8\pi D^2}
\left\langle
\Lambda_{ij,kl}(\hat{\vb*{n}})
\Lambda_{ij,mn}(\hat{\vb*{n}})
\dddot I_{kl}\dddot I_{mn}
\right\rangle\,,
\label{T0rretardedfluxdetail}
\end{equation}
where Eq.~\eqref{hTTfarzonefluxdetail} has been used in the last equality.
Then, we obtain
\begin{equation}
\mathcal{F}_{\rm T}
=
\frac{G_N}{8\pi}
\int {\rm d}\Omega\,
\left\langle
\Lambda_{ij,kl}
\Lambda_{ij,mn}
\dddot I_{kl}\dddot I_{mn}
\right\rangle\,.
\label{Fgangularintegraldetail}
\end{equation}
The TT projectors satisfy the angular-averaging 
relation
\begin{equation}
\int {\rm d}\Omega\,\Lambda_{ij,kl}\Lambda_{ij,mn}
=\frac{8\pi}{5}\left[
\frac12(\delta_{km}\delta_{ln}+\delta_{kn}\delta_{lm})
-\frac13\delta_{kl}\delta_{mn}
\right]\,.
\label{TTaverageidentityfluxdetail}
\end{equation}
Using Eq.~(\ref{TTaverageidentityfluxdetail}), 
with all spatial indices understood as those 
of the original source-frame coordinates, the total
radiated power in Eq.~(\ref{Fgangularintegraldetail}) becomes
\begin{equation}
\mathcal{F}_{\rm T}
=\frac{G_N}{5}
\left\langle 
\dddot I_{ij}\dddot I_{ij}
- \frac13 (\dddot I_{kk})^2 
\right\rangle.
\label{Fgstartdetail}
\end{equation}

For a binary in the center-of-mass frame, 
the mass quadrupole moment is given by Eq.~(\ref{Iijpp}), namely $I_{ij}=m_{r} x_i x_j$, where
$\vb*{x}=\vb*{r}=(x,y,z)$. 
The nonvanishing components of $I_{ij}$ are
listed in Eq.~(\ref{Ixyz}). 
Since $I_{kk}=I_{xx}+I_{yy}=m_{r} r^2$ is
constant for circular motion, we have $\dddot I_{kk}=0$ in
Eq.~(\ref{Fgstartdetail}). In addition,
$\dddot I_{ij}\dddot I_{ij}
=(\dddot I_{xx})^2+(\dddot I_{yy})^2
+2(\dddot I_{xy})^2
=32m_{r}^2 r^4\omega^6$.
Therefore, 
\begin{equation}
\mathcal{F}_{\rm T} = \frac{32}{5}G_N m_{r}^2 r^4\omega^6
=\frac{32}{5}\frac{G_N}{G_{\rm eff}^2}\eta^2 v^{10}\,,
\label{Fgrwdetail}
\end{equation}
where, in the second equality, we used 
Eq.~(\ref{vrel1}) and $m_{r}=\eta M$.
This result has the same functional dependence 
as in GR, but the overall prefactor 
$G_N/G_{\rm eff}^2$ is kept explicit, while 
the relation among $v$, $r$, and $\omega$ is modified through $G_{\rm eff}$.

\subsection{Scalar dipole flux}
\label{Scafsec}

To derive the scalar dipole flux, we consider the propagation of the scalar
perturbation $\varphi$. In the far (wave) zone, we neglect the contribution
of the term $\mu_{,\phi}(\phi)F$ in Eq.~(\ref{phieom2}), since it is quadratic
in the vector field and therefore subleading compared to the matter source
term in the weak-field regime. 
Approximating $\sqrt{-g} \simeq 1$ and
$u_I^{0} \simeq 1$, the radiative field $\varphi$ obeys
\begin{equation}
\Box \varphi(x)=\frac{\rho_s(x)}{M_{\rm Pl}}\,,
\label{waveeqphifluxdetail}
\end{equation}
where, in the nonrelativistic limit, the scalar source density is given by
\begin{equation}
\rho_s(t,\bm{x}) 
\simeq \sum_I \alpha_I m_I 
\delta^{(3)}\!\bigl(\bm{x}-\bm{x}_I(t)\bigr)\,.
\end{equation}
At the observer position $(t,\bm{X})$, 
the retarded solution of 
Eq.~(\ref{waveeqphifluxdetail}) is
\begin{equation}
\varphi(t,\bm{X})
=-\frac{1}{4\pi M_{\rm Pl}}
\int {\rm d}^3x'\,
\frac{\rho_s\!\bigl(t-|\bm{X}-\bm{x}'|,\bm{x}'\bigr)}
{|\bm{X}-\bm{x}'|}\,,
\label{retphifluxdetail}
\end{equation}
where $\bm{x}'$ is the integration variable 
representing the source position.
In the far zone, with $D=|\bm{X}| \gg |\bm{x}'|$, 
we expand $|\bm{X}-\bm{x}'|$ as in Eq.~(\ref{bXx}). 
Using $\hat{\bm n}= \bm{X}/D$ and the 
retarded time $t_r=t-D$, 
we approximate the denominator in Eq.~(\ref{retphifluxdetail}) by its 
leading term $D$. Corrections to this denominator are suppressed by powers of 
$|\bm{x}'|/D$ and hence do not affect the leading radiative field. On the other hand, the term linear in $\bm{x}'$ 
is kept in the retarded time, 
as it determines the phase of the outgoing radiation.
Then,
\begin{equation}
\varphi(t,\bm{X})
\simeq
-\frac{1}{4\pi M_{\rm Pl} D}
\int {\rm d}^3x'\,
\rho_s\!\left( t_r+\hat{\bm n}\cdot \bm{x}',\bm{x}'\right).
\label{varphiin}
\end{equation}
Expanding the source around the retarded time $t_r$ gives
\begin{equation}
\rho_s\!\left(t_r+\hat{\bm n}\cdot \bm{x}',\bm{x}'\right)
=\rho_s(t_r,\bm{x}')
+(\hat{\bm n}\cdot \bm{x}')\,\partial_{t_r} \rho_s(t_r,\bm{x}')
+\cdots.
\end{equation}
Substituting this into the above expression yields
\begin{equation}
\varphi(t,\bm{X})
=-\frac{1}{4\pi M_{\rm Pl} D}
\left[ q_s(t_r)+\hat n_i \dot{\mathcal D}_\phi^{\,i}(t_r)+\cdots
\right],
\label{varphisoX}
\end{equation}
where
\begin{equation}
q_s(t_r) \equiv \int {\rm d}^3x'\,\rho_s(t_r,\bm{x}'),
\qquad
\mathcal D_\phi^{\,i}(t_r)\equiv \int {\rm d}^3x'\,x'^i \rho_s(t_r,\bm{x}').
\label{qsD}
\end{equation}
Here, a dot denotes a derivative with respect to $t_r$.

For point particles, the quantities (\ref{qsD}) reduce to
\begin{equation}
q_s=\sum_I \alpha_I m_I,
\qquad
\bm{\mathcal D}_\phi=\sum_I \alpha_I m_I \bm{x}_I,
\end{equation}
which correspond to the scalar charge and scalar 
dipole moment, respectively.
Since the scalar charges $\alpha_I m_I$
of the individual compact objects are constant at leading order, the
total scalar monopole charge $q_s$ is conserved. 
Hence, the monopole term does not radiate, and the leading radiative contribution is
\begin{equation}
\varphi(t,\bm{X})
=-\frac{1}{4\pi M_{\rm Pl} D}\,
\hat{n}_i\dot{\mathcal D}_{\phi}^{\,i}(t_r).
\label{phiraddetail}
\end{equation}
In the far zone, the energy-momentum tensor 
of the scalar perturbation is given by
\begin{equation}
T_{\mu\nu}^{(\varphi)}
=\partial_\mu \varphi\,\partial_\nu \varphi
-\frac12 \eta_{\mu\nu}(\partial\varphi)^2\,.
\end{equation}
For outgoing waves, the scalar field depends on the retarded time $t-D$, so that $\partial_D\varphi=-\partial_t\varphi$. Hence, the radial energy-flux
component is
\begin{equation}
T_{0D}^{(\varphi)}
=-(\partial_t\varphi)^2
=-\frac{1}{16\pi^2 M_{\rm Pl}^2 D^2}
\,\hat{n}_i\hat{n}_j
\ddot{\mathcal D}_{\phi}^{\,i}\ddot{\mathcal D}_{\phi}^{\,j}.
\end{equation}
The scalar radiated power is given by 
\begin{equation}
{\cal F}_{\rm S}
= -D^2 \int {\rm d}\Omega\, T_{0D}^{(\varphi)}\,.
\end{equation}
Using 
$\int {\rm d}\Omega\,\hat{n}_i\hat{n}_j=(4\pi/3)\delta_{ij}$, 
we obtain
\begin{equation}
\mathcal{F}_{\rm S}
=\frac{1}{12\pi M_{\rm Pl}^2}
\left\langle \ddot{\bm{\mathcal D}}_{\phi}^{\,2}\right\rangle
=\frac{2}{3}G_N
\left\langle \ddot{\bm{\mathcal D}}_{\phi}^{\,2}\right\rangle.
\label{Fphi}
\end{equation}

For a binary system composed of two compact objects 
labelled by $A$ and $B$, the scalar dipole moment is 
$\bm{\mathcal D}_\phi = 
\alpha_A m_A \bm{x}_A+\alpha_B m_B \bm{x}_B$.
Using the center-of-mass relations 
$\bm{x}_A=m_B \bm{r}/M$ and $\bm{x}_B=-m_A \bm{r}/M$, we obtain 
\begin{equation}
\bm{\mathcal D}_{\phi}=m_{r} \Delta \alpha\,\bm{r},
\label{Dphi}
\end{equation}
where $\Delta \alpha \equiv \alpha_A-\alpha_B$.
Since $\ddot{\bm r}=-\omega^2 \bm r$ for circular motion, 
it follows that $\ddot{\bm{\mathcal D}}_{\phi}^{\,2}=m_{r}^2 (\Delta \alpha)^2 r^2 \omega^4$.
Substituting this result into Eq.~(\ref{Fphi}), 
we obtain
\begin{equation}
\mathcal{F}_{\rm S}
= \frac{2}{3}G_N m_{r}^2 (\Delta \alpha)^2 
r^2\omega^4
= \frac{2G_N}{3 G_{\rm eff}^2}\eta^2 (\Delta \alpha)^2 v^{8}.
\label{Fsrwdetail}
\end{equation}
Compared with the tensor radiated power in Eq.~(\ref{Fgrwdetail}), which is
proportional to $v^{10}$, the scalar contribution 
scales as $v^8$ and therefore enters at $-1$PN order.

\subsection{Vector dipole flux}
\label{Vecfsec}

In the vector sector, we solve the wave equation for $A^\mu$ using the retarded Green's function. 
In the far zone ($r \to \infty$), the scalar field approaches 
a constant value, $\phi \to \phi_\infty$, so that the coupling 
function becomes $\mu(\phi) \to \mu_\infty = \mathrm{const}$. 
Choosing the Lorenz gauge condition, $\partial_\mu A^\mu=0$, 
Eq.~(\ref{Fmunueq}) reduces to 
\begin{equation}
\Box A^\mu (x) = \frac{1}{\mu_\infty} 
J_A^\mu(x)\,.
\label{BoxA}
\end{equation}
In the weak-field limit, the source current (\ref{JAmu}) is approximated as 
\be
J_{A}^{\mu} (t, {\bm x}) \simeq
\sum_{I} q_I u_I^\mu\,
\delta^{(3)}(\bm{x}-\bm{x}_I(t))\,.
\label{JAmu2}
\ee
The retarded solution to Eq.~(\ref{BoxA}) at the observer position $(t,{\bm X})$ is given by 
\begin{equation}
A^{\mu}(t,\bm{X}) 
= -\frac{1}{4\pi \mu_\infty} 
\int {\rm d}^3x' 
\frac{
J_A^{\mu}\bigl(t-|\bm{X}-\bm{x}'|,\bm{x}'\bigr)}
{|\bm{X}-\bm{x}'|}\,.
\label{Amuso}
\end{equation}
Expanding $|\bm{X}-\bm{x}'|$ at large distances as in Eq.~(\ref{bXx}), the charge density $\rho_e = J_A^{0}$ 
can be expressed as
\begin{equation}
\rho_e(t-|\bm{X}-\bm{x}'|,\bm{x}')
= \rho_e(t_r,\bm{x}')
+ (\hat{\bm n}\cdot \bm{x}')\, \partial_{t_r} \rho_e(t_r,\bm{x}')
+ \cdots\,.
\end{equation}
The monopole contribution is proportional to the electric charge $q=\int {\rm d}^3 x\,\rho_e = \sum_I q_I$, which is conserved and hence does not generate radiation.
The leading nonvanishing radiative contribution 
arises from the dipole term,
which involves the first moment of the charge distribution.
For a system of point particles, the electric dipole moment 
is thus given by 
\begin{equation}
\bm{\mathcal D}_{e}=\int {\rm d}^3x\, 
\rho_e(t,\bm{x})\,\bm{x}
= \sum_I q_I \bm{x}_I\,.
\label{DAes}
\end{equation}
Using the charge parametrization introduced in Eq.~(\ref{sigmaDefApp}) 
for a binary labelled by $A$ and $B$, 
Eq.~(\ref{DAes}) can be expressed as 
\be
\bm{\mathcal D}_{e}=\sqrt{4\pi\mu_\infty G_N}\,
m_{r}\,\Delta \sigma\,\bm{r}\,,
\label{DAdetail2}
\ee
where 
\be
\Delta \sigma \equiv \sigma_A-\sigma_B\,.
\ee
Here, we have used the fact that the coupling function 
$\mu(\phi)$ approaches the common asymptotic value 
$\mu_\infty$ outside both bodies.

We derive the vector radiated power $\mathcal{F}_{\rm V}$ from the
energy-momentum tensor of the vector field. 
In the far zone, where $\mu(\phi)\to\mu_\infty$, 
this tensor is given by
\begin{equation}
T^{(A)}_{\mu\nu}
=\mu_\infty \left(
F_{\mu\alpha}F_{\nu}{}^{\alpha}
-\frac14 \eta_{\mu\nu}F_{\alpha\beta}F^{\alpha\beta}
\right)\,.
\label{TmunuAdetail}
\end{equation}
In terms of the electric and magnetic fields, ${\bm E}$ and ${\bm B}$, this
tensor gives the usual Poynting vector multiplied by $\mu_\infty$. 
With our sign convention, the radial energy-flux component is
\begin{equation}
T_{0D}^{(A)}
=-\mu_\infty (\bm{E}\times \bm{B})\cdot \hat{\bm n}\,.
\label{T0rAdetail}
\end{equation}
Using the radiation-zone relations
$\bm{B}=\hat{\bm n}\times \bm{E}$ and
$\hat{\bm n}\cdot \bm{E}=0$, we find
$\bm{E}\times \bm{B}=\bm{E}^2\hat{\bm n}$. Hence,
\begin{equation}
T_{0D}^{(A)}=-\mu_\infty \bm{E}^{2}\,.
\label{T0rAraddetail}
\end{equation}
The vector radiated power is therefore
\begin{equation}
\mathcal{F}_{\rm V}
=-D^2 \int {\rm d}\Omega\,T_{0D}^{(A)}
=\mu_\infty D^2 \int {\rm d}\Omega\, \bm{E}^{2}\,.
\label{FAfromTmunudetail}
\end{equation}
In analogy with the derivation of 
Eq.~(\ref{phiraddetail}) from
Eq.~(\ref{varphiin}) in the scalar sector, 
the temporal component of
Eq.~(\ref{Amuso}) gives the far-zone 
dipole contribution
\be
A^0(t,\bm{X}) \simeq
-\frac{1}{4\pi\mu_\infty D}\,
\hat{n}_i \dot{\mathcal D}_{e}^{i}(t_r)\,,
\label{A0so0}
\ee
where $\mathcal D_{e}^{i}$ denotes the $i$-th 
component of $\bm{\mathcal D}_{e}$. 
The spatial component of Eq.~(\ref{Amuso}) 
in the far zone is
\be
A^i (t,\bm{X})
\simeq -\frac{1}{4\pi\mu_\infty D}
\int {\rm d}^3 x'\, J_A^i(t_r,\bm{x}')\,.
\label{Airad}
\ee
Taking the time derivative of 
$\mathcal D_e^i= \int {\rm d}^3x' \rho_e x'^i$, 
using the continuity equation $\partial_t \rho_e + \partial_j J_A^j = 0$, 
and performing integration by parts, we find
\begin{equation}
\dot{\mathcal D}_e^i
=\int {\rm d}^3x'\, J_A^i\,.
\end{equation}
Therefore, Eq.~(\ref{Airad}) can be written as
\be
A^i (t,\bm{X})
=-\frac{1}{4\pi\mu_\infty D}
\dot{\mathcal D}_e^i(t_r)\,.
\label{Aiso}
\ee

The $j$-th component of the radiative electric field is given by 
\be
E_j= -\partial_j A^0 - \partial_t A_j\,.
\label{Ej}
\ee
Using $\partial_j t_r=-\partial_j D=-X_j/D=-\hat n_j$, 
and noting that the dominant contribution to $\partial_j A^0$ 
arises from the dependence of the source on the 
retarded time, we obtain
\be
\partial_j A^0
\simeq
-\frac{1}{4\pi\mu_\infty D}
\hat{n}_i \ddot{\mathcal D}_{e}^{i}\,\partial_j t_r
=\frac{1}{4\pi\mu_\infty D}
\hat{n}_i \hat{n}_j \ddot{\mathcal D}_{e}^{i}\,.
\ee
Since $\partial_t A_j=-\delta_{ij}\ddot{\mathcal D}_e^i/(4\pi \mu_{\infty}D)$, 
Eq.~(\ref{Ej}) reduces to 
\be
E_j
=\frac{1}{4\pi\mu_\infty D}P_{ij}\ddot{\mathcal D}_e^{\,i}\,,
\label{Ejdef}
\ee
where $P_{ij}$ is the projection tensor defined in Eq.~(\ref{Lambda}).
Using the relation
$P_{ij}P_{ik}=\delta_{jk}-\hat{n}_j\hat{n}_k$, 
we find
\be
\bm{E}^{2}
=\frac{1}{16\pi^2\mu_\infty^2 D^2}
\left[
\ddot{\bm{\mathcal D}}_e^{\,2}
-(\hat{\bm n}\cdot \ddot{\bm{\mathcal D}}_e)^2
\right]\,.
\label{Erad2detail}
\ee
Substituting Eq.~\eqref{Erad2detail} into Eq.~\eqref{FAfromTmunudetail}, 
the radiated power becomes
\begin{equation}
\mathcal{F}_{\rm V}
=\frac{1}{16\pi^2\mu_\infty}
\int {\rm d}\Omega
\left[
\ddot{\bm{\mathcal D}}_e^{\,2}
-(\hat{\bm n}\cdot \ddot{\bm{\mathcal D}}_e)^2
\right]\,.
\end{equation}
Using the angular averages $\int {\rm d}\Omega = 4\pi$
and $\int {\rm d}\Omega\,\hat{n}_i\hat{n}_j
=(4\pi/3)\delta_{ij}$, we arrive at
\be
\mathcal{F}_{\rm V}
=\frac{1}{6\pi\mu_\infty}
\left\langle \ddot{\bm{\mathcal D}}_e^{\,2}\right\rangle\,.
\label{FAstartdetail}
\ee
From Eq.~\eqref{DAdetail2}, we have
$\ddot{\bm{\mathcal D}}_e^{\,2}=
4\pi\mu_\infty G_N\,m_{r}^2(\Delta\sigma)^2 r^2\omega^4$, 
so that 
\be
\mathcal{F}_{\rm V}
= \frac{2G_N}{3}m_{r}^2(\Delta\sigma)^2 r^2\omega^4
=\frac{2G_N}{3G_{\rm eff}^2}\eta^2(\Delta\sigma)^2 v^8\,.
\label{FArwdetail}
\ee
This contribution also enters at $-1$PN order relative to the tensor
quadrupole flux.

\subsection{Total flux and frequency sweep}
\label{Totalfsec}

In the preceding subsections, we derived the energy fluxes associated with the tensor, scalar, and vector radiation channels. We now combine these contributions to obtain the total energy flux emitted by the binary system.
Summing Eqs.~\eqref{Fgrwdetail}, \eqref{Fsrwdetail}, 
and \eqref{FArwdetail}, we obtain
\begin{align}
\mathcal{F}_{\rm tot}
&=\mathcal{F}_{\rm T}+\mathcal{F}_{\rm S}+\mathcal{F}_{\rm V}
\nonumber\\
&= \frac{32}{5} \frac{G_N}{G_{\rm eff}^2}\eta^2 v^{10}
\left( 1 + b v^{-2}\right)\,,
\label{Ftotal2}
\end{align}
where 
\be
b \equiv \frac{5}{48} \left[(\Delta\alpha)^2+(\Delta\sigma)^2\right]\,.
\label{bdef}
\ee
Here, $\Delta\alpha$ and $\Delta\sigma$ denote the differences in the scalar
and electric charge-to-mass ratios of the two bodies, respectively. The parameter $b$ therefore measures the mismatch in these charge-to-mass ratios
and vanishes for binaries with identical scalar and electric charge-to-mass ratios. The factor $v^{-2}$ 
shows that the dipole contributions enter at
$-1$PN order relative to the tensor quadrupole flux and can dominate in the early inspiral when $\Delta\alpha$ or $\Delta\sigma$ is nonzero. In ESM
theories, the scalar charge is not a primary quantity, but can be induced by
the electric charge through the coupling $\mu(\phi)F$. Thus, a nonvanishing
$\Delta\sigma$ can also generate a nonzero $\Delta\alpha$, depending on the
form of the coupling and the background configuration.

For a binary system with orbital energy $E$, 
the radiation discussed above
leads to an energy loss governed by
\begin{equation}
\dv{E}{t} = -\mathcal{F}_{\rm tot}
=-\frac{32}{5} \frac{G_N}{G_{\rm eff}^2}\eta^2 v^{10}
\left( 1+b v^{-2} \right)\,.
\label{balance}
\end{equation}
For a circular orbit, the Newtonian binding 
energy is given by Eq.~\eqref{Eene}. 
Taking the derivative of
$E(\omega)=-m_{r}(G_{\rm eff} M\omega)^{2/3}/2$
with respect to $\omega$, and combining the result with
Eq.~\eqref{balance}, we obtain the evolution of the orbital frequency,
\begin{equation}
\dot{\omega} = \frac{96}{5} G_N 
\left(G_{\rm eff}^{2}M_c^{5}\omega^{11}\right)^{1/3}
\left[1+b\,(G_{\rm eff}M\omega)^{-2/3}\right]\,,
\label{omegadot}
\end{equation}
where we used $\eta=m_{r}/M$, $M_c=\eta^{3/5}M$, and
$v=(G_{\rm eff}M\omega)^{1/3}$. The standard gravitational radiation result
in GR is recovered in the limit $b \to 0$ and $G_{\rm eff} \to G_N$, for which
$\dot{\omega}>0$. For $b>0$, the scalar and vector dipole losses further
accelerate the inspiral.
The effective gravitational coupling,
$G_{\rm eff}=G_N (1+2\alpha_A \alpha_B-\sigma_A \sigma_B)$, also modifies the frequency sweep.
However, in the inspiral regime with $v\ll 1$, 
the correction proportional to $b v^{-2}$ 
gives the leading deviation from the GR result, provided that $b v^{-2}$ remains smaller than unity. 
The modification due solely to
$G_{\rm eff}$ is instead of relative order
$\alpha_A\alpha_B$ or $\sigma_A\sigma_B$, 
and hence does not contain the
dipole enhancement factor $v^{-2}$.

In terms of the binary relative velocity $v=(G_{\rm eff} M\omega)^{1/3}$,
Eq.~\eqref{omegadot} can be expressed as
\begin{equation}
\dot v = \frac{32}{5}\frac{G_N}{G_{\rm eff}^2}
\frac{\eta}{M}v^9\left(1+bv^{-2}\right)\,.
\label{vdot}
\end{equation}
This equation can be inverted to obtain a first-order differential equation
for $t$ with respect to $v$. 
Assuming $b v^{-2}\ll 1$, we expand
$(1 + bv^{-2})^{-1}$ to first order in $b v^{-2}$ and obtain
\begin{equation}
t_c-t \simeq \frac{5M}{256\eta}\frac{G_{\rm eff}^2}
{G_N}v^{-8} \left(1-\frac{4}{5}bv^{-2}\right)\,,
\label{tcminustv}
\end{equation}
where $t_c$ is the coalescence time of the binary.

The emission of radiation drives the slow evolution of the orbital frequency and hence modifies the accumulated orbital phase. 
Using ${\rm d}\Phi/{\rm d}v=\omega\,{\rm d}t/{\rm d}v$ with $\omega=v^3/(G_{\rm eff}M)$, the integrated solution for $\Phi$
up to linear order in $bv^{-2}$ is given by
\be
\Phi_c-\Phi (v)=\frac{1}{32\eta} \frac{G_{\rm eff}}{G_N}
v^{-5}\left(1-\frac{5}{7}bv^{-2}\right)\,,
\label{Phiv}
\ee
where $\Phi_c$ is the orbital phase at coalescence. 

In Sec.~\ref{Fouriersec}, we will exploit the relations 
\eqref{tcminustv} and \eqref{Phiv} to compute the gravitational waveform 
in the frequency domain.

\subsection{Orbital-period decay}
\label{orbital}

Observations of binary pulsars place stringent constraints on additional
radiation channels from NSs. 
In particular, the measured orbital-period
decay of the Hulse-Taylor pulsar PSR B1913+16
\cite{Hulse:1974eb,Taylor:1982zz} is in excellent agreement with the
tensor-quadrupole prediction of GR, thereby severely limiting possible
contributions from dipole radiation.

We derive the dipole parameter $\kappa_D$ associated with the intrinsic orbital-period decay of a compact binary in ESM theories. Since both the scalar
and vector fields are massless, dipole radiation contributes at leading order
during the inspiral. For a circular binary orbit, the orbital separation $r$ is related to the orbital velocity $v$ by $r=G_{\rm eff}M/v^2$. The orbital
period is then given by
\begin{equation}
P_b=\frac{2\pi r}{v}=2\pi G_{\rm eff} M v^{-3}\,.
\label{PbkappaApp}
\end{equation}
Taking the time derivative of 
Eq.~(\ref{PbkappaApp}) and using
Eq.~(\ref{vdot}), we obtain
\begin{equation}
\dot{P}_b
=-\frac{192\pi}{5}\frac{G_N}{G_{\rm eff}}\eta v^5
\left(1+bv^{-2}\right)\,.
\label{PbdotkappaApp}
\end{equation}
Since radiation reaction increases $v$, 
the orbital period decreases with time. 
The tensor-quadrupole contribution, denoted by
$\dot{P}_b^{\rm quad}$, corresponds to the limit 
$b \to 0$, namely
\begin{equation}
\dot{P}_b^{\rm GR}
=-\frac{192\pi}{5}\frac{G_N}{G_{\rm eff}}\eta v^5\,.
\label{PbdotGRkappaApp}
\end{equation}
Using $\dot{P}_b^{\rm GR}$, Eq.~(\ref{PbdotkappaApp}) can be rewritten as
\begin{equation}
\dot{P}_b
=\dot{P}_b^{\rm GR}\left(1+\kappa_D v^{-2}\right)\,,
\label{PbkappaDefApp}
\end{equation}
where the dipole parameter $\kappa_D$ is given by 
\begin{equation}
\kappa_D=b=
\frac{5}{48}\left[(\Delta\alpha)^2+(\Delta\sigma)^2\right]\,.
\end{equation}
Thus, the dipole correction is determined by the differences in the scalar
and vector charge-to-mass ratios of the 
two bodies, and the two contributions
enter additively at leading order.

The observational constraint on the orbital-period decay places direct bounds on dipole radiation from NSs. 
After subtracting kinematic effects, the observed
orbital decay rate is consistent with the GR prediction as
\cite{Weisberg:2016jye}
\begin{equation}
\frac{\dot{P}_b^{\rm obs}}{\dot{P}_b^{\rm GR}}
= 0.9983 \pm 0.0016 \,.
\end{equation}
In our parametrization (\ref{PbkappaDefApp}), 
this bound translates to
\begin{equation}
\kappa_D v^{-2} = -0.0017 \pm 0.0016 \,.
\end{equation}
For the Hulse-Taylor binary pulsar PSR B1913+16, 
the typical orbital velocity is 
$v \simeq 1.5 \times 10^{-3}$, 
which leads to the constraint
\begin{equation}
b \lesssim 10^{-8} \,.
\label{bbound}
\end{equation}
This bound applies to NS-NS binaries, but not 
directly to BH-BH or ECO-ECO binaries. 
For the latter systems, constraints on $b$ 
can instead be obtained
from observations of inspiral gravitational waveforms.

\section{Frequency-domain tensor waveform}
\label{Fouriersec}

In this section, we derive the frequency-domain gravitational waveform by
taking into account the evolution of the orbital 
frequency driven by radiation reaction. 
This is the form most directly used in matched-filter
searches and parameter estimation. 
As shown in Sec.~\ref{Fresec}, the emission of tensor, scalar, and vector radiation from the binary leads to a
secular increase in the orbital frequency, characterized by the frequency sweep $\dot{\omega}$. 
We focus on the leading tensor quadrupole contribution and the 
leading scalar and vector dipole corrections, 
neglecting higher-PN terms.
Working at this order, we incorporate the time dependence of
$\omega(t)$ and construct the frequency-domain waveform using the stationary phase approximation (SPA).

\subsection{Waveforms under SPA}

As shown in Eq.~(\ref{hp}), the plus polarization of GWs
in the time domain is given by
\begin{equation}
h_+(t)= -\mathcal{A}(t)\cos[2\Phi(t)],
\qquad
\mathcal{A}(t)=\frac{4G_N}{D}M_c^{5/3}(G_{\rm eff} \omega)^{2/3}
\frac{1+\cos^2\iota}{2}\,,
\label{hplusSPAstart}
\end{equation}
where $\Phi(t)$ denotes the orbital phase evaluated at the retarded time
$t-D$, so that $\dot{\Phi}(t)=\omega(t)$ up to an irrelevant constant
shift in the time origin. The term proportional to $D$ contributes only
a constant phase shift to $\Phi(t)$ and does not affect the stationary
phase condition. Hence, it can be absorbed into the overall phase
without loss of generality.

Using $\cos(2\Phi)=(e^{2i\Phi}+e^{-2i\Phi})/2$ 
and adopting the Fourier transform convention
$\tilde h(f)=\int_{-\infty}^{\infty}{\rm d}t\,h(t)
e^{-2\pi i f t}$, the Fourier transform of $h_+(t)$ can be written as
\be
\tilde h_+(f)
=\int_{-\infty}^{\infty} {\rm d}t\, h_+(t)e^{-2\pi i f t} 
=-\frac{1}{2}\int_{-\infty}^{\infty} {\rm d}t\, \mathcal{A}(t)
\left\{ e^{-i[2\pi f t-2\Phi(t)]}+e^{-i[2\pi f t+2\Phi(t)]}\right\}\,.
\label{htrans}
\ee
The integral contains two rapidly oscillating phases,
$2\pi f t-2\Phi(t)$ and $2\pi f t+2\Phi(t)$.
For $f>0$, the contribution from the phase $2\pi f t+2\Phi(t)$
does not admit a stationary point and is therefore highly suppressed. 
By contrast, the term with phase
\begin{equation}
\Psi_f(t)\equiv 2\pi f t-2\Phi(t)
\end{equation}
does admit a stationary point. 
The stationary time $t_f$ is determined by the condition
$\dot{\Psi}_f(t_f)=0$. Using 
$\dot{\Phi}(t_f)=\omega(t_f)$, this condition gives
\begin{equation}
\omega(t_f)=\pi f\,.
\label{stationary}
\end{equation}
Defining
\begin{equation}
u_f \equiv (\pi G_{\rm eff} M f)^{1/3}\,,
\label{nudef}
\end{equation}
we have $u_f=v(t_f)$, where we used the relation 
$\omega=v^3/(G_{\rm eff} M)$.

To evaluate the dominant contribution, we expand both the phase and the slowly varying amplitude around $t=t_f$. Since the amplitude varies on the radiation-reaction timescale, 
whereas the phase oscillates on the orbital
timescale, it is sufficient at leading SPA order to replace
$\mathcal{A}(t)$ by $\mathcal{A}(t_f)$ and expand 
the phase up to quadratic order:
\begin{equation}
\Psi_f(t)=\Psi_f(t_f)+\frac12 
\ddot{\Psi}_f(t_f)(t-t_f)^2+\cdots.
\label{Psiexpand}
\end{equation}
Then, the Fourier integral reduces to a Gaussian integral, 
\be
\tilde h_+(f)
\simeq 
-\frac{1}{2} \int {\rm d}t\, 
\mathcal{A}(t) e^{-i \Psi_f(t)} 
\simeq
-\frac12 \mathcal{A}(t_f)e^{-i\Psi_f(t_f)}
\int_{-\infty}^{\infty} {\rm d}t\,
\exp\!\left[-\frac{i}{2}\ddot{\Psi}_f(t_f)(t-t_f)^2\right] \,.
\label{SPAplus0a}
\ee
Using
$\int_{-\infty}^{\infty}{\rm d}t\,
\exp\!\left( -ia t^2/2\right)
=\sqrt{2\pi/|a|}\,
e^{-i\,{\rm sign}(a)\pi/4}$,
we obtain
\begin{equation}
\tilde h_+(f)
\simeq
-\frac12 \mathcal{A}(t_f)
\sqrt{\frac{2\pi}{|\ddot{\Psi}_f(t_f)|}}
\,e^{-i[\Psi_f(t_f)-\pi/4]}\,.
\label{SPAplus0}
\end{equation}
Since $\ddot{\Psi}_f(t_f)=-2\dot{\omega}(t_f)<0$ during the inspiral, the
Gaussian integral gives the phase factor $e^{i\pi/4}$.
Hence,
Eq.~(\ref{SPAplus0}) can be written as 
\begin{equation}
\tilde h_+(f)
\simeq
-\frac12 \mathcal{A}(t_f)
\sqrt{\frac{\pi}{\dot{\omega}(t_f)}}
\,e^{-i\Psi_+(f)}\,,
\qquad
\Psi_+(f)\equiv 2\pi f t_f-2\Phi(t_f)-\frac{\pi}{4}.
\label{SPAplus1}
\end{equation}
Using Eq.~(\ref{omegadot}) with $\omega(t_f)=\pi f$, 
and expanding to linear order in $b u_f^{-2}$, we finally obtain
\be
\tilde h_+(f)
=
-\sqrt{\frac{5}{24}}\,
\frac{\sqrt{G_N}\,G_{\rm eff}^{1/3}M_c^{5/6}}
{\pi^{2/3} D}
\frac{1+\cos^2\iota}{2}
\,f^{-7/6}
\left(1-\frac12 b u_f^{-2}\right)e^{-i\Psi_+(f)}\,.
\label{hplusF}
\ee
Similarly, from Eq.~(\ref{hm}),
\be
\tilde h_\times(f)
=-\sqrt{\frac{5}{24}}\,
\frac{\sqrt{G_N}\,G_{\rm eff}^{1/3}M_c^{5/6}}
{\pi^{2/3} D}(\cos\iota)
\,f^{-7/6}
\left(1-\frac12 b u_f^{-2}\right)e^{-i\Psi_\times(f)}\,,
\label{hcrossF}
\ee
where
\begin{equation}
\Psi_\times(f)=\Psi_+(f)+\frac{\pi}{2}\,.
\end{equation}
This phase shift reflects the $\pi/2$ difference between the two
tensor polarizations. In the GR limit $G_{\rm eff}\to G_N$ and $b\to 0$,
these expressions reduce to the standard leading-order 
Fourier amplitudes.

The SPA phase in Eq.~(\ref{SPAplus1}) can be written as  
\begin{equation}
\Psi_+(f)=2\pi f t_c-2\Phi_c-\frac{\pi}{4}
-2\pi f (t_c-t_f)+2\left[\Phi_c-\Phi(t_f)\right]\,,
\end{equation}
where $t_c$ and $\Phi_c$ denote the coalescence time 
and the orbital phase at coalescence, respectively. 
Using Eqs.~(\ref{tcminustv}) and (\ref{Phiv}), 
we have
\begin{equation}
2\pi f (t_c-t_f)=\frac{5}{128\eta}
\frac{G_{\rm eff}}{G_{N}} 
u_f^{-5}\left(1-\frac{4}{5}b u_f^{-2}\right),
\qquad
\Phi_c-\Phi(t_f)
=
\frac{1}{32\eta}\frac{G_{\rm eff}}{G_N}
u_f^{-5}\left(1-\frac{5}{7}b u_f^{-2}\right),
\end{equation}
where we used $v(t_f)=u_f$ and  
$2\pi f=2\omega(t_f)=2u_f^3/(G_{\rm eff}M)$.  
Substituting these expressions into $\Psi_+(f)$, we obtain
\begin{equation}
\Psi_+(f)=2\pi f t_c - 2\Phi_c - \frac{\pi}{4}
+ \frac{3}{128\eta}\frac{G_{\rm eff}}{G_N}u_f^{-5}
\left(1-\frac{4}{7} b u_f^{-2}\right).
\label{psioff}
\end{equation}
More explicitly, this can be written as
\begin{equation}
\Psi_+(f)=2\pi f t_c - 2\Phi_c - \frac{\pi}{4}
+ \frac{3}{128\eta}\frac{G_{\rm eff}}{G_N}(\pi G_{\rm eff} M f)^{-5/3}
-\frac{5}{3584\eta}\frac{G_{\rm eff}}{G_N}
\left[(\Delta\alpha)^2+(\Delta\sigma)^2\right](\pi G_{\rm eff}M f)^{-7/3}.
\label{psioff2}
\end{equation}
The last term represents the leading dipole 
correction to the phase. 
It enters at $-1$PN order, as expected 
for dipole radiation.

\subsection{GWs on a cosmological background}

GWs emitted from a compact binary can propagate over cosmological distances before reaching the observer. 
It is therefore necessary to take into account
the effects of the cosmological background on the waveform, such as the redshift of the frequency and the amplitude dilution 
due to cosmic expansion. In what follows, we derive the frequency-domain waveform including these
cosmological effects.

Let us consider a spatially flat cosmological background,
whose line element is given by
\begin{equation}
{\rm d}s^2=-{\rm d}t^2+a^2(t)\delta_{ij}{\rm d}x^i{\rm d}x^j\,,
\end{equation}
where the scale factor $a(t)$ depends on the cosmic time $t$.
We denote by $t_s$ and $t_0$ the times of emission and observation, respectively. 
The redshift of the source is then defined as
\begin{equation}
1+z=\frac{a(t_0)}{a(t_s)}\,.
\end{equation}
The frequency measured by the observer, $f_0$, 
is related to the source-frame frequency, $f_s$, as
\begin{equation}
f_0=(1+z)^{-1} f_s\,.
\label{freqredshift}
\end{equation}
Correspondingly, the angular frequency satisfies
$\omega_0=(1+z)^{-1}\omega_s$. In the gravitational waveform, the phase
evolution depends on the velocity variable
$(G_{\rm eff} M \omega)^{1/3}$. Hence, when expressed in terms of the observed
frequency $f_0$, this combination can be written in terms of the redshifted
masses
\begin{equation}
{\cal M} \equiv (1+z)M\,,\qquad
{\cal M}_c \equiv (1+z)M_c\,,
\label{redshiftedmass}
\end{equation}
which are the mass parameters directly 
inferred from observations. 
In particular, ${\cal M}_c$ is the redshifted chirp mass. 
The velocity variable entering the waveform can then be expressed in terms of observer-frame quantities as
\begin{equation}
u \equiv (\pi G_{\rm eff}{\cal M} f_0)^{1/3}
= (\pi G_{\rm eff} M f_s)^{1/3}\,.
\label{u0def}
\end{equation}
Thus, $u$ coincides with the source-frame orbital 
velocity evaluated at the stationary point.

In Minkowski spacetime, the waveform amplitudes in
Eqs.~(\ref{hplusF}) and (\ref{hcrossF}) scale as 
$1/D$, where $D$ is the physical distance to the source. 
In a cosmological background, this distance is identified with the physical distance at the time of observation,
$D\to a(t_0)r$, where $r$ is the comoving distance. 
Using the definition of
the luminosity distance, $d_L(z)=(1+z)a(t_0)r$, we have
\begin{equation}
D \to a(t_0)r = (1+z)^{-1}d_L(z)\,.
\end{equation}
The Fourier-domain waveform is also affected by the time dilation between
the source and observer frames, ${\rm d}t_0=(1+z){\rm d}t_s$, together with
the frequency redshift $f_s=(1+z)f_0$. 
These factors combine with the redshifted chirp mass
${\cal M}_c=(1+z)M_c$, so that the overall 
GW amplitude takes the standard
form proportional to ${\cal M}_c^{5/6}f_0^{-7/6}/d_L(z)$.

In the present ESM theories, tensor modes 
undergo the standard cosmological
damping determined by the scale factor, because the Einstein-Hilbert term has
a constant Planck mass and there is no nonminimal coupling of the form ${\cal F}(\phi)R$. Hence, the above result directly applies, with the luminosity
distance $d_L(z)$ entering the waveform amplitude. In more general theories with a time-dependent effective Planck mass, the luminosity distance should
instead be replaced by the GW distance $d_{\rm GW}(z)$
\cite{Saltas:2014dha,Lombriser:2015sxa,
Belgacem:2017ihm,Belgacem:2019pkk,Nishizawa:2017nef,Quartin:2023tpl}.

The observed frequency-domain tensor polarizations 
are then obtained from Eqs.~(\ref{hplusF}) and (\ref{hcrossF}) by consistently incorporating
the redshift of the frequency, the time dilation 
in the Fourier transform, and the relation between the physical and luminosity distances discussed above.
This leads to
\begin{align}
\tilde h_+(f_0)
&=
-\sqrt{\frac{5}{24}}\,
\frac{\sqrt{G_N}\,G_{\rm eff}^{1/3}{\cal M}_c^{5/6}}
{\pi^{2/3} d_L(z)}
\frac{1+\cos^2\iota}{2}
\,f_0^{-7/6}
\left(1-\frac12 b u^{-2}\right)e^{-i\Psi_+(f_0)}\,,
\label{hplusFcosmo}
\\
\tilde h_\times(f_0)
&=
-\sqrt{\frac{5}{24}}\,
\frac{\sqrt{G_N}\,G_{\rm eff}^{1/3}{\cal M}_c^{5/6}}
{\pi^{2/3} d_L(z)}
(\cos\iota)
\,f_0^{-7/6}
\left(1-\frac12 b u^{-2}\right)e^{-i\Psi_\times(f_0)}\,,
\label{hcrossFcosmo}
\end{align}
where the observer-frame SPA phase of the plus polarization is
\begin{equation}
\Psi_+ (f_0)=2\pi f_0 t_{c,0}-2\Phi_c-\frac{\pi}{4}
+\frac{3}{128\eta}\frac{G_{\rm eff}}{G_N}u^{-5}
\left(1-\frac{4}{7}b u^{-2}\right)\,,
\label{psicosmo}
\end{equation}
and
\begin{equation}
\Psi_\times(f_0)=\Psi_+(f_0)+\frac{\pi}{2}\,.
\end{equation}
Here, $t_{c,0}$ is the coalescence time in the observer 
frame, while $\Phi_c$ is the orbital phase at coalescence. Equivalently, using Eqs.~(\ref{bdef}) and (\ref{u0def}), we can write
\begin{equation}
\Psi_+(f_0)
=2\pi f_0 t_{c,0}-2\Phi_c-\frac{\pi}{4}
+\frac{3}{128\eta}\frac{G_{\rm eff}}{G_N}
(\pi G_{\rm eff}{\cal M} f_0)^{-5/3}
-\frac{5}{3584\eta}\frac{G_{\rm eff}}{G_N}
\left[(\Delta\alpha)^2+(\Delta\sigma)^2\right]
(\pi G_{\rm eff}{\cal M} f_0)^{-7/3}\,.
\label{psicosmo2}
\end{equation}
Equations~(\ref{hplusFcosmo})-(\ref{psicosmo2}) provide the
frequency-domain tensor waveforms in a cosmological background.
The cosmological redshift is absorbed into the redshifted masses
${\cal M}$ and ${\cal M}_c$ and the observed frequency $f_0$, while the amplitude is suppressed by the luminosity distance $d_L(z)$.

It is useful to rewrite the above result in a form analogous to the
parameterized post-Einsteinian (ppE) 
waveform \cite{Yunes:2009ke,Cornish:2011ys,Tahura:2018aea} in a cosmological
background. Using the redshifted masses ${\cal M}$ and ${\cal M}_c$
introduced above, together with the luminosity distance $d_L(z)$,
we define the GR reference waveform for the plus polarization as
\begin{equation}
\tilde h_{\rm GR}(f_0)
=-{\cal A}_{\rm GR}\, f_0^{-7/6} 
e^{-i\Psi_{{\rm GR},+}(f_0)}\,,
\end{equation}
where 
\ba
{\cal A}_{\rm GR}
&=& 
\sqrt{\frac{5}{24}}\,
\frac{(G_N {\cal M}_c)^{5/6}}
{\pi^{2/3} d_L(z)}\,,
\\
\Psi_{{\rm GR},+}(f_0)
&=& 
2\pi f_0 t_{c,0}-2\Phi_c-\frac{\pi}{4}
+\frac{3}{128\eta}
(\pi G_N{\cal M} f_0)^{-5/3}\,.
\ea
We also introduce
\begin{equation}
u_N \equiv (\pi G_N{\cal M} f_0)^{1/3},
\qquad
u \equiv (\pi G_{\rm eff}{\cal M} f_0)^{1/3}
=
\left(\frac{G_{\rm eff}}{G_N}\right)^{1/3}u_N .
\end{equation}
The conservative modification associated with $G_{\rm eff}\neq G_N$
can be separated from the dipole-radiation correction by defining
\begin{equation}
\Delta\Psi_{\rm eff}
\equiv
\frac{3}{128\eta}
\left(
\frac{G_{\rm eff}}{G_N}u^{-5}-u_N^{-5}
\right)
=
\frac{3}{128\eta}u_N^{-5}
\left[
\left(\frac{G_{\rm eff}}{G_N}\right)^{-2/3}-1
\right].
\end{equation}
The two tensor polarizations (\ref{hplusFcosmo}) and 
(\ref{hcrossFcosmo}) can then be written as
\begin{align}
\tilde h_+(f_0)
&=
\tilde h_{\rm GR}(f_0)
\left(\frac{G_{\rm eff}}{G_N}\right)^{1/3}
\frac{1+\cos^2\iota}{2}
\left( 1+\alpha_{\rm ppE} u^{-2}\right)
\exp\!\left[ i\beta_{\rm ppE} u^{-7}
-i\Delta\Psi_{\rm eff} \right],
\\
\tilde h_\times(f_0)
&=
\tilde h_{\rm GR}(f_0)
\left(\frac{G_{\rm eff}}{G_N}\right)^{1/3}
(\cos\iota)
\left( 1+\alpha_{\rm ppE} u^{-2}\right)
\exp\!\left[ i\left(\beta_{\rm ppE} u^{-7}-\frac{\pi}{2}\right) 
-i\Delta\Psi_{\rm eff}\right],
\end{align}
where the ppE parameters associated with the dipole radiation are
given by
\ba
\alpha_{\rm ppE}
&=& -\frac{1}{2}\,b
= -\frac{5}{96}\left[(\Delta\alpha)^2
+(\Delta\sigma)^2\right]\,,
\\
\beta_{\rm ppE}
&=& \frac{3}{224\eta}\frac{G_{\rm eff}}{G_N}\,b
= \frac{5}{3584\,\eta}\frac{G_{\rm eff}}{G_N}
\left[(\Delta\alpha)^2+(\Delta\sigma)^2\right].
\ea
Nonvanishing values of $\Delta\alpha$ and $\Delta\sigma$ give rise to
scalar and vector dipole radiation, thereby inducing the leading
$-1$PN corrections to the amplitude and phase through the ppE
parameters $\alpha_{\rm ppE}$ and $\beta_{\rm ppE}$. Since the phase
correction $\beta_{\rm ppE}u^{-7}$ enters at $-1$PN order, it is
enhanced at low frequencies and can dominate the inspiral phase
modification over the conservative contribution 
$\Delta\Psi_{\rm eff}$.
Note that $\Delta\Psi_{\rm eff}$ originates from the conservative
modification of the binary dynamics through $G_{\rm eff}$.

\section{Application to BH-BH, ECO-ECO, and NS-NS binaries}
\label{Secapp}

In the preceding sections, we showed that scalar and vector dipole
radiation gives rise to leading $-1$PN corrections to the inspiral
waveform through the parameter $b$. These corrections affect both the
amplitude and the phase, while the phase contribution is enhanced in
the low-frequency inspiral regime. Hence, GW observations can tightly
constrain the parameter $b$. In this section, we relate 
this waveform parameter to the scalar and vector charges 
carried by concrete compact objects, including BHs, ECOs, and NSs.

The dimensionless scalar charge parameter $\alpha_I$ 
is related to the
scalar charge $q_{s_I}$ of the compact object labeled by $I$. We define
$q_{s_I}$ through the asymptotic behavior of the scalar field at large
distances from the object,
\be
\phi(r)=\phi_{\infty}-\frac{q_{s_I}}{r}\,.
\ee
As discussed in Refs.~\cite{Damour:1992we,Higashino:2022izi}, 
the derivative of the ADM mass $m_I(\phi)$ with respect to $\phi$ 
can be expressed in terms of the scalar charge as
$m_{I,\phi}=4\pi q_{s_I}$. Using the definition of $\alpha_I$, we then obtain
\be
\alpha_I=\frac{4\pi M_{\rm Pl} q_{s_I}}{m_I}\,.
\ee
This relation is consistent with Eq.~(\ref{varphixI}). 
We also recall that the dimensionless vector charge-to-mass ratio $\sigma_I$ is
defined in Eq.~(\ref{sigmaDefApp}).

For a binary system consisting of two compact objects with masses
$m_A$ and $m_B$, vector charges $q_A$ and $q_B$, and scalar charges
$q_{s_A}$ and $q_{s_B}$, the parameter
$b=(5/48)[(\Delta \alpha)^2+(\Delta \sigma)^2]$ can be written as
\be
b=\frac{5 M_{\rm Pl}^2}{24\,\mu_{\infty}} 
\left[
\left( \frac{q_A}{m_A}-\frac{q_B}{m_B} \right)^2
+8\pi^2 \mu_{\infty} 
\left( \frac{q_{s_A}}{m_A}-\frac{q_{s_B}}{m_B} \right)^2
\right].
\label{bana}
\ee
Thus, $b$ is controlled by the differences in the vector and scalar
charge-to-mass ratios of the two compact objects. In particular, the
dipole contribution vanishes when both charge-to-mass ratios are the
same for the two bodies.

The magnitudes of the scalar charges $q_{s_A}$ and $q_{s_B}$ acquired
by compact objects depend on the form of the coupling $\mu(\phi)$ and
on the internal structure of each object. 
In the following, we consider
three representative cases:
(A) BH-BH binaries with the exponential coupling
$\mu(\phi)=e^{-\gamma \phi/M_{\rm Pl}}$,
(B) ECO-ECO binaries with $\mu(\phi)\propto \phi^{-p}$ ($p>0$) near
the center, and
(C) NS-NS binaries with the coupling
$\mu(\phi)=e^{-\beta \phi^2/M_{\rm Pl}^2}$.

\subsection{BH-BH binaries with exponential coupling}

For BH-BH binary systems, we consider the exponential coupling
\be
\mu(\phi)=e^{-\gamma \phi/M_{\rm Pl}}\,,
\label{exponential}
\ee
where $\gamma$ is a dimensionless constant. 
In the convention of the action (\ref{action}), 
and with the normalization $M_{\rm Pl}^2=2$, the dilatonic coupling corresponding to heterotic string theory
\cite{Gibbons:1987ps,Garfinkle:1990qj} is 
obtained for $\gamma^2=2$.
The RN BH solution is recovered in the limit $\gamma\to 0$. 
In ESM theories, BHs are linearly stable against both 
odd- and even-parity perturbations, provided that $\mu(\phi)>0$
\cite{Gannouji:2021oqz,Kase:2023kvq}. This stability condition is
automatically satisfied by the exponential 
coupling (\ref{exponential}).

To derive electrically charged BH solutions with scalar hair, it is
convenient to adopt the static and spherically symmetric line element
in the form
\be
{\rm d}s^2=-f(\hat{r})\,{\rm d}t^2+f^{-1}(\hat{r})\,{\rm d}\hat{r}^2
+r^2(\hat{r})\left({\rm d}\theta^2+\sin^2\theta\,{\rm d}\varphi^2\right)\,,
\ee
where both $f$ and $r$ are functions of $\hat{r}$. 
On this background, the vector field $A_\mu$ has only a nonvanishing temporal component, $A_0(\hat{r})$.

\subsubsection{$\gamma^2 \neq 2$} 

For $\gamma^2 \neq 2$, there exists the following 
asymptotically flat exact solution:
\ba
& &
f(\hat{r})=\left( 1-\frac{r_+}{\hat{r}}\right)
\left( 1-\frac{r_{-}}{\hat{r}}\right)^{\frac{2-\gamma^2}{2+\gamma^2}}\,,
\qquad
r(\hat{r})=\hat{r}\left( 1-\frac{r_{-}}{\hat{r}}
\right)^{\frac{\gamma^2}{2+\gamma^2}}\,,
\nonumber\\
& &
A_0'(\hat{r})=\frac{q\, e^{\gamma \phi_{\infty}/M_{\rm Pl}}}{\hat{r}^2}\,,
\qquad 
\phi(\hat{r})=\phi_{\infty}+\frac{2\gamma}
{2+\gamma^2}M_{\rm Pl}\ln\!\left( 
1-\frac{r_{-}}{\hat{r}} \right)\,,
\label{GMso}
\ea
where $r_+$, $r_{-}$, $q$, and $\phi_{\infty}$ are integration
constants. The parameter $q$ corresponds to 
the electric charge and obeys
\be
q^2=\frac{4M_{\rm Pl}^2\, r_+ r_{-}\, 
e^{-\gamma \phi_{\infty}/M_{\rm Pl}}}{2+\gamma^2}\,.
\label{qre}
\ee
This solution describes a BH with a curvature singularity at
$\hat{r}=r_{-}$, which corresponds to $r=0$. The event horizon is
located at $\hat{r}=r_+$, and 
the curvature singularity is hidden by
the horizon for $r_+>r_-$. For simplicity, we set
$\phi_{\infty}=0$ in what follows, so that $\mu_{\infty}=1$.

Expanding $r(\hat{r})$ at large $\hat{r}$, we obtain
\be
r(\hat{r})
=
\hat{r}
-\frac{\gamma^2}{\gamma^2+2}r_-
+{\cal O}(\hat r^{-1})\,.
\ee
Thus, $\hat{r}$ coincides with the areal radius $r$ at leading order in the large-distance limit. 
The ADM mass $m$ of the BH can be
identified from the asymptotic expansion of $f(\hat{r})$,
\be
f(\hat{r})=1-\frac{2G_N m}{\hat{r}}
+{\cal O}(\hat{r}^{-2})\,,
\ee
which yields
\be
m=4\pi \Mpl^2 \left( r_{+}+\frac{2-\gamma^2}
{2+\gamma^2}r_{-} \right)\,.
\label{mre}
\ee
The large-distance expansion of $\phi(\hat{r})$ 
in Eq.~(\ref{GMso})
gives the scalar charge
\be
q_s=\frac{2\gamma \Mpl r_{-}}{2+\gamma^2}\,.
\label{qs}
\ee
Using Eqs.~(\ref{qre}) and (\ref{mre}), we obtain 
\be
r_{-}=\frac{m(2+\gamma^2)}{8\pi \Mpl^2 (2-\gamma^2)}
\left[ 1-\sqrt{1-\frac{16\pi^2 \Mpl^2 q^2 (2-\gamma^2)}
{m^2}} \right]\,,
\ee
where we have selected the branch that recovers
$r_{-}\to 0$ in the limit $q\to 0$. Substituting this into
Eq.~(\ref{qs}), we obtain
\be
q_s=\frac{m \gamma}{4\pi \Mpl(2-\gamma^2)}\left[ 
1-\sqrt{1-\frac{16\pi^2 \Mpl^2 q^2 (2-\gamma^2)}{m^2}} \right]\,.
\label{qsdef}
\ee
Since $q_s$ vanishes in the limit $q\to 0$, the scalar charge
represents a secondary hair induced by the primary electric charge $q$.

We consider a BH-BH binary system whose components 
have electric charges $q_I$ and masses $m_I$, 
where $I=A, B$. 
We define the dimensionless charge parameter
\be
\hat{q}_I \equiv \frac{q_I}{\Mpl r_{h_I}}\,,
\qquad 
r_{h_I}=2G_N m_I\,.
\label{hatqI}
\ee
Then, the quantity $b$ defined in Eq.~(\ref{bana}) becomes
\be
b=\frac{5 (\hat{q}_A - \hat{q}_B)^2}{384 \pi^2} 
\left[ 1+\frac{8 \pi^2 \gamma^2 (\hat{q}_A+ \hat{q}_B)^2}
{(\sqrt{1 - (2-\gamma^2 )\hat{q}_A^2}+\sqrt{1 
- (2-\gamma^2 )\hat{q}_B^2})^2} \right]\,.
\label{bgene}
\ee
For $\hat{q}_A \neq \hat{q}_B$, the parameter $b$ 
is nonzero. Thus, a difference in the charge-to-mass ratios of the two BHs sources dipolar radiation, leading to modifications of both the amplitude and
phase of GWs. Even when $q_A=q_B$, a mass 
asymmetry $m_A\neq m_B$ generically gives 
$\hat q_A\neq \hat q_B$, or equivalently
$q_A/m_A\neq q_B/m_B$.

\subsubsection{$\gamma^2=2$}

The case $\gamma^2=2$ corresponds to 
the BH solutions found by 
Gibbons and Maeda~\cite{Gibbons:1987ps} and 
Garfinkle, Horowitz, and Strominger~\cite{Garfinkle:1990qj}. 
These solutions can be obtained as the 
$\gamma^2\to 2$ limit of
Eqs.~(\ref{GMso}) and (\ref{qre}). In the following, we choose
$\gamma=\sqrt{2}$ and set $\phi_{\infty}=0$. Then Eq.~(\ref{qre}) gives
\begin{equation}
r_{+}=\frac{q^2}{M_{\rm Pl}^2 r_{-}}\,.
\end{equation}
Using the relation $m=4\pi M_{\rm Pl}^2 r_{+}$, 
which follows from
Eq.~(\ref{mre}) for $\gamma^2=2$, 
we can write $r_{-}$ as
\begin{equation}
r_{-}=\frac{4\pi q^2}{m}\,.
\end{equation}
Then, the scalar charge in Eq.~(\ref{qs}) reduces to
\begin{equation}
q_s=\frac{2\sqrt{2}\,\pi M_{\rm Pl}\, q^2}{m}\,.
\end{equation}
For $\gamma=-\sqrt{2}$, the sign of $q_s$ is reversed, while the expression for $b$ given below is unchanged. Defining $\hat{q}_I$ as
in Eq.~(\ref{hatqI}) for the binary, 
Eq.~(\ref{bana}) yields
\begin{equation}
b=\frac{5 (\hat{q}_A - \hat{q}_B)^2}{384 \pi^2} 
\left[ 1+4\pi^2 (\hat{q}_A+ \hat{q}_B)^2 \right],
\label{baspe}
\end{equation}
so that $b \neq 0$ for $\hat{q}_A \neq \hat{q}_B$. 
The second term in the square brackets of Eq.~(\ref{baspe}) is recovered by taking the limit $\gamma^2 \to 2$ of the corresponding term in Eq.~(\ref{bgene}).

\subsubsection{$\gamma=0$}

The special case $\gamma=0$ corresponds to 
the RN BH solution. 
For $\phi_{\infty}=0$, we have $r=\hat r$, 
$A_0'(r)=q/r^2$, and $\phi(r)=0$, 
so that the scalar charge vanishes.
The metric function is given by
\begin{equation}
f(r)=\left(1-\frac{r_{+}}{r}\right)
\left(1-\frac{r_{-}}{r}\right),
\end{equation}
where
\begin{equation}
r_{+}=\frac{q^2}{2 M_{\rm Pl}^2 r_{-}}, \qquad 
r_{-}=\frac{m}{8\pi M_{\rm Pl}^2}  
\left[1-\sqrt{1-\frac{32\pi^2 M_{\rm Pl}^2 q^2}{m^2}}\right],
\end{equation}
represent the outer and inner horizons, respectively. 
The branch for $r_-$ has been chosen such 
that $r_- \to 0$ in the neutral limit 
$q \to 0$, while $r_+\to 2G_N m$.
For $\gamma=0$, the scalar charge
vanishes, $q_s=0$, as follows from Eq.~(\ref{qsdef}). 
Using Eq.~(\ref{bgene}), the parameter $b$ 
for a binary system reduces to
\begin{equation}
b=\frac{5(\hat{q}_A-\hat{q}_B)^2}{384\pi^2}.
\label{bRN}
\end{equation}
In contrast to the case $\gamma \neq 0$, 
there is no scalar-induced
contribution proportional to $(\hat{q}_A+\hat{q}_B)^2$.

\subsection{ECO-ECO binaries}

Let us now turn to the case of ECO-ECO binaries.
For couplings that diverge near the center of a compact object, 
it is possible to realize ECO solutions without 
a curvature singularity at $r=0$ 
\cite{Herdeiro:2019mbz,DeFelice:2024ops,DeFelice:2025vef}. 
In such models, the coupling function $\mu(\phi)$ behaves near the center as
$\mu(\phi) \propto (\phi - \phi_0)^{-p}$, where $\phi_0$ is the
scalar-field value at the center and $2 < p \le 3$
\cite{DeFelice:2026cse}. 
The divergence of $\mu(\phi)$ at $r=0$ corresponds to a weak effective
coupling near the center, while all physical quantities remain finite. 
Unlike NSs, these ECOs can be supported solely by the scalar-vector 
interaction $\mu(\phi)F$, without introducing additional matter sources
such as a perfect fluid. 

We consider a static and spherically symmetric background described by
the line element
\be
{\rm d}s^2=-f(r){\rm d}t^2+h^{-1}(r) {\rm d}r^2
+r^2 \left( {\rm d}\theta^2+\sin^2 \theta\,
{\rm d}\varphi^2 \right)\,,
\label{line}
\ee
where $f$ and $h$ are functions of $r$.
For a general coupling $\mu(\phi)$, the vector-field equation gives
\be
A_0'(r)=\frac{q\sqrt{N}}{\mu r^2}\,,
\label{A0so}
\ee
where $q$ is an integration constant corresponding to the vector charge, and $N=f/h$. 
We focus on asymptotically flat solutions satisfying
$f,h \to 1$ as $r \to \infty$, and set $\mu_\infty=1$.
In this large-distance limit, Eq.~(\ref{A0so}) 
reduces to $A_0'(r) \simeq q/r^2$.
The equation of motion for the scalar 
field $\phi(r)$ is
\be
\left[ \sqrt{N}\, r^2 h\,\phi'(r) \right]'
=-\frac{r^2 \mu_{,\phi}}{2\sqrt{N}}A_0'(r)^2\,,
\label{phieq}
\ee
where a prime denotes differentiation 
with respect to $r$.
For regular ECOs, we impose the boundary condition 
$\sqrt{N}\, r^2 h\, \phi'(r) \to 0$ as $r \to 0$.
The scalar charge $q_s$ is defined by 
the asymptotic limit
$q_s=\lim_{r \to \infty} r^2 \phi'(r)$.
Substituting Eq.~(\ref{A0so}) into Eq.~(\ref{phieq}) and integrating from the center to spatial infinity, we obtain
\be
q_s=-\frac{q^2}{2} \int_{0}^{\infty} 
\frac{\mu_{,\phi}\sqrt{N}}{\mu^2 r^2}\,{\rm d}r\,.
\label{qsint}
\ee
For a given coupling function $\mu(\phi)$, the scalar charge can be evaluated from Eq.~(\ref{qsint}).
The ADM mass is given by 
$m=\lim_{r \to \infty} 4\pi M_{\rm Pl}^2\, r \left[1-h(r)\right]$.
For an ECO-ECO binary system characterized by the vector charges $q_I$ and masses $m_I$ ($I=A,B$), the parameter $b$ in Eq.~(\ref{bana}) can then be computed.

For the model in which the function $N(r)$ takes the form
\cite{DeFelice:2024ops}
\be
N(r)=\left( \frac{r^4+\sqrt{N_0} r_0^4}
{r^4+r_0^4} \right)^2\,,
\label{Nana}
\ee
where $N_0$ and $r_0$ are positive constants, the coupling function 
$\mu(\phi)$ can be reconstructed so as to realize the analytic 
expression (\ref{Nana}) \cite{DeFelice:2025vef}.
In this case, the coupling exhibits the behavior
$\mu(\phi) \propto (\phi-\phi_0)^{-3}$ near $r=0$, whereas at spatial
infinity it approaches $\mu_\infty$ as
$\mu(\phi)-\mu_\infty \propto \phi-\phi_\infty$. 
At large distances, the scalar-field derivative behaves as
$\phi'(r) \propto r^{-3}$, which implies $q_s=0$.
Defining $\hat{q}_I$ as in Eq.~(\ref{hatqI}) for the ECO-ECO binary,
we then obtain
\be
b=\frac{5 (\hat{q}_A - \hat{q}_B)^2}{384 \pi^2}\,,
\ee
which has the same form as Eq.~(\ref{bRN}).
This demonstrates that an ECO-ECO binary in the analytic model
(\ref{Nana}) modifies the gravitational waveform in a manner analogous
to a binary composed of RN BHs.

Another model that realizes ECOs with regular centers is described by
the coupling~\cite{DeFelice:2026cse}
\be
\mu(\phi)=\mu_0+\mu_1 \frac{M_{\rm Pl}^p}{(\phi-\phi_0)^p}\,,
\label{mup}
\ee
where $\mu_0$ and $\mu_1$ are positive constants, 
with $2 < p \le 3$. 
In this model, the large-distance behavior of $\phi'(r)$ is given by
$\phi'(r) \propto r^{-2}$ \cite{DeFelice:2026cse}, implying that the
ECO carries a nonvanishing scalar charge.
To compute $q_s$ explicitly, one needs to 
numerically integrate the right-hand side of Eq.~(\ref{qsint}). 
For an ECO-ECO binary system described by 
the coupling (\ref{mup}), both $q_{s_A}$ and 
$q_{s_B}$ are generally nonzero in Eq.~(\ref{bana}).
They therefore give rise to additional waveform modifications sourced by the scalar charges, compared with the analytic model~(\ref{Nana}).

\subsection{NS-NS binaries}

Finally, we consider binary systems composed of 
NSs endowed with both vector charges $q_I$ 
and scalar charges $q_{s_I}$. Such NS solutions
are expected to arise for coupling functions $\mu(\phi)$ of the type given in 
Eqs.~(\ref{exponential}) and~(\ref{mup}). 
For couplings
$\mu(\phi)$ containing even powers of $\phi$, spontaneous scalarization
can occur for charged BHs
\cite{Herdeiro:2018qqt,Fernandes:2019rez,Ikeda:2019qjp}
and NSs \cite{Minamitsuji:2021vdb} in strong-gravity environments,
in analogy with the standard spontaneous scalarization mechanism induced
by a nonminimal scalar coupling to 
the Ricci scalar
\cite{Damour:1993hw}.
As an illustrative example, consider the coupling
\be
\mu(\phi)=e^{-\beta \phi^2/M_{\rm Pl}^2}\,,
\label{muspo}
\ee
which admits two distinct branches of solutions:
(i) the GR branch, characterized by $\phi(r)=0$ everywhere, and
(ii) the scalarized branch, for which $\phi(r)\neq 0$.
For a negative coupling constant $\beta$, the GR branch can become
unstable against a tachyonic instability as the ratio of the charge
density $\rho_c$ to the matter density $\rho_m$ increases, thereby
driving the system toward the scalarized branch.

On the static and spherically symmetric background described by the
line element (\ref{line}), the baryonic matter inside the NS can be
modeled as a perfect fluid with radially dependent density
$\rho_m(r)$ and pressure $P_m(r)$. 
We also introduce a charge density
$\rho_c(r)$. 
The temporal component of the vector field satisfies
\be
\left[ \mu r^2 \frac{A_0'(r)}{\sqrt{N}} 
\right]'=\frac{\rho_c(r) r^2}{\sqrt{h}}\,,
\label{A0eq}
\ee
where $N=f/h$. Imposing the regular boundary condition
$\mu r^2 A_0'(r)/\sqrt{N} \to 0$ as $r \to 0$, we obtain
\be
A_0'(r)=\frac{\sqrt{N}}{\mu\,r^2} \int_0^r 
\frac{\rho_c(\tilde{r})\,\tilde{r}^2}{\sqrt{h(\tilde{r})}} 
\,{\rm d}\tilde{r}\,.
\ee
The charge density $\rho_c$ is nonvanishing only inside the star of
radius $r_s$. The total vector charge is then given by
\be
q= \int_0^{r_s} 
\frac{\rho_c(\tilde{r})\,\tilde{r}^2}{\sqrt{h(\tilde{r})}} 
\,{\rm d}\tilde{r}\,.
\ee
Outside the stellar surface ($r>r_s$), the electric field takes the
form $A_0'(r)=q\sqrt{N}/(\mu r^2)$, as in Eq.~(\ref{A0so}).

The scalar field obeys an equation of the same form as
Eq.~(\ref{phieq}). Provided that
$\sqrt{N} r^2 h \phi'(r) \to 0$ as $r \to 0$, the derivative
$\phi'(r)$ can be expressed in the integrated form
\be
\phi'(r)=-\frac{1}{\sqrt{N}r^2 h} 
\int_0^r \frac{\tilde{r}^2 \mu_{,\phi}}{2\sqrt{N(\tilde{r})}}
A_0'(\tilde{r})^2 \,{\rm d}\tilde{r}\,.
\ee
As long as both $h$ and $N$ approach unity at spatial infinity, the
scalar charge is given by
\begin{equation}
q_s=-\int_0^{\infty}
\frac{\tilde{r}^2 \mu_{,\phi}}{2\sqrt{N(\tilde{r})}}
A_0'(\tilde{r})^2 \,{\rm d}\tilde{r}\,.
\end{equation}
To evaluate this quantity, one must integrate the remaining background
equations for the metric functions, together with
\begin{equation}
P_m'+\frac{f'}{2f}(\rho_m+P_m)=\frac{\rho_c A_0'}{\sqrt{f}} \,,
\end{equation}
starting from the vicinity of $r=0$. The stellar radius is determined
by the condition $P_m(r_s)=0$. 
In Ref.~\cite{Minamitsuji:2021vdb}, the radius 
and ADM mass of scalarized NS solutions were 
computed numerically for the
coupling (\ref{muspo}) with a specified NS 
equation of state, under the
assumption that the ratio $\rho_c/\rho_m$ remains constant inside the star.
In this way, the value of $b$ for a NS-NS binary can be evaluated
numerically once the quantities $q$, $q_s$, and $m$ are determined for
each star.

For a NS-NS binary system, the Hulse-Taylor binary pulsar PSR B1913+16
provides the bound (\ref{bbound}) on the parameter $b$ from the
observed orbital-period decay. Setting $\mu_{\infty}=1$ in
Eq.~(\ref{bana}), this bound translates into
\begin{equation}
\frac{5 M_{\rm Pl}^2}{24} 
\left[ \left( \frac{q_A}{m_A}-\frac{q_B}{m_B} \right)^2
+8\pi^2 
\left( \frac{q_{s_A}}{m_A}-\frac{q_{s_B}}{m_B} 
\right)^2 \right] \lesssim 10^{-8} \,.
\label{binarybo}
\end{equation}
This implies the approximate upper limits
\begin{equation}
M_{\rm Pl}\left|\frac{q_A}{m_A}-\frac{q_B}{m_B}\right|
\lesssim 2\times 10^{-4},
\qquad
M_{\rm Pl}\left|\frac{q_{s_A}}{m_A}
-\frac{q_{s_B}}{m_B}\right|
\lesssim 2\times 10^{-5},
\end{equation}
for NS-NS binaries with vector and scalar charges. 
Note that the bound (\ref{binarybo}) does not apply to 
binary systems composed of BHs or ECOs. 
In such cases, GW observations can instead place 
constraints on $b$ through the waveform.

\section{Conclusions and future outlook}
\label{consec}

In this paper, we have investigated inspiral gravitational waveforms
from compact binaries in ESM theories, with particular emphasis on the
role of scalar and vector charges in both the conservative dynamics and
the radiative sector. Our analysis consistently connects the near-zone
dynamics, the far-zone energy flux, and the frequency-domain tensor
waveform, allowing us to identify the leading deviations from GR in a
systematic manner.

In Sec.~\ref{SecNear}, we formulated ESM theories 
and derived the near-zone solutions governing the conservative dynamics of compact
binaries. By solving the quasi-static field equations in the weak-field
regime, we obtained the gravitational, scalar, and vector potentials
generated by point-like sources. 
This allowed us to express the binary
dynamics in terms of an effective gravitational coupling $G_{\rm eff}$, which incorporates 
the effects of scalar and vector interactions. 
In the subsection on circular orbits, we applied these results to a binary system on a circular orbit and derived the relation between the
orbital separation and angular frequency. This provides the basis for
connecting the conservative dynamics to the waveform calculations 
in the subsequent sections.

In Sec.~\ref{SecTen}, we derived the tensor gravitational waveform in
the time domain. Working in the far (wave) zone, we solved the
linearized Einstein equations using the retarded Green's function under
the harmonic gauge condition. We then projected the metric perturbations
onto the TT components relevant for GW observations. The resulting
tensor waveform retains the same angular dependence as in GR, reflecting
the tensorial nature of gravitational radiation, while its amplitude is
modified through the appearance of $G_{\rm eff}$ in the orbital
dynamics. This demonstrates explicitly how modifications in the
near-zone dynamics propagate into observable waveform amplitudes.

In Sec.~\ref{Fresec}, we computed the energy loss due to radiation in the tensor, scalar, and vector sectors. By evaluating the energy fluxes
in the far zone, we showed that scalar and vector dipole radiation
enters at $-1$PN order relative to the leading tensor quadrupole
emission. These contributions are encoded in the parameter $b$, which
depends on the differences in scalar and vector charge-to-mass ratios
between the two bodies. We derived the modified energy balance equation
and demonstrated how the presence of dipole radiation enhances the
inspiral rate compared to GR.\@ This result has direct implications for
the orbital-period decay of compact binaries and provides a key link to
observational constraints from binary pulsars.

In Sec.~\ref{Fouriersec}, we constructed the frequency-domain gravitational waveform using the stationary phase approximation. Starting from the time-domain signal, we derived analytic expressions
for the Fourier amplitude and phase that incorporate the modified frequency evolution. 
We showed that dipole radiation affects both the
amplitude and the phase, with the latter receiving a characteristic
$-1$PN correction proportional to $b$. This correction accumulates over
many cycles and is therefore particularly sensitive to GW observations.
We also extended the waveform to a cosmological background, incorporating
the effects of redshift and the luminosity distance on the amplitude,
frequency, and phase evolution. 
Finally, we presented a mapping onto the
parameterized post-Einsteinian framework, which facilitates model-independent tests of gravity 
against observational data.

In Sec.~\ref{Secapp}, we applied the general 
waveform results to concrete compact-object binaries, including BH-BH, ECO-ECO, and NS-NS
systems. For BH-BH binaries, we considered electrically charged solutions with scalar hair in the exponential coupling model and
derived the corresponding expression of the dipole parameter $b$ in terms of the charge-to-mass ratios. 
For ECO-ECO binaries, we showed that regular-center solutions supported by the scalar-vector interaction can carry vector charges and, depending on the coupling, scalar charges as well. 
In particular, the analytic ECO model with $q_s=0$
gives a waveform modification analogous to that 
of RN BH binaries, whereas more general regular ECO models can yield additional
scalar-charge contributions. For NS-NS binaries, we discussed charged
stars with scalar hair, including the possibility of spontaneous
scalarization, and related the parameter $b$ to the vector and scalar
charges of the two stars. We also showed how binary-pulsar measurements
of the orbital-period decay provide stringent bounds on the differences
in the charge-to-mass ratios.

In the Appendix, we derived the scalar and vector radiation fields and
obtained their waveforms in both the time and frequency domains. Unlike
in nonminimally coupled scalar-tensor theories, the scalar and vector
sectors in ESM theories do not mix with the tensor sector at the linear
level. As a result, they do not generate additional metric
polarizations beyond the two tensor modes. However, their emission
modifies the orbital evolution through energy loss, thereby affecting
the tensor waveform indirectly, in particular through phase corrections.

The framework developed in this work opens several directions for future
investigation. Including higher PN corrections would improve the
accuracy of waveform templates and allow for more precise parameter
estimation. Incorporating spin-orbit and spin-spin couplings of compact
objects would also be important for realistic modeling, since these
effects influence the phase evolution and are routinely included in GW
data analyses.

From an observational perspective, combining GW measurements with
binary-pulsar timing provides a powerful probe of deviations from GR.
The parameter $b$ can be tightly constrained by current and future
observations, especially in systems involving NSs. For NSs and ECOs,
tidal deformations are generally nonvanishing
\cite{Flanagan:2007ix,Hinderer:2007mb}, in contrast to BHs in GR.\@
Such finite-size effects leave characteristic imprints on the GW signal
\cite{LIGOScientific:2017vwq,De:2018uhw,Diedrichs:2025vhv}, and their
inclusion can help break degeneracies among parameters, thereby allowing
for tighter constraints on $b$ and other deviations from GR.

Looking ahead, next-generation detectors, both ground-based and
space-based, will further improve the sensitivity to dipole-radiation
effects. It is also interesting to explore possible imprints of scalar
and vector radiation in stochastic GW backgrounds and pulsar timing
arrays, where cumulative effects over many sources may become
detectable. These multi-band and complementary observations will provide
a promising avenue to test ESM theories and further constrain
deviations from GR.

\section*{Acknowledgements}

S.T. acknowledges support from JSPS KAKENHI Grant Nos.~26K07090 
and 26H00847, and from the Waseda University Special Research 
Projects (No.~2025C-488).

\appendix

\section*{Appendix:~Scalar and vector waves}
\label{AppenA}

In this Appendix, we derive the scalar and vector waveforms in both
the time and frequency domains in ESM theories. 
A key difference from
nonminimally coupled scalar-tensor theories 
is that the ESM
action~\eqref{action} does not include a coupling between the scalar
field and the Ricci scalar of the form 
${\cal F}(\phi)R$. As a result, the
metric perturbation sourced by a compact binary exhibits only the usual
transverse-traceless tensor polarizations at linear order, while the
scalar and vector fields propagate as independent radiative degrees of
freedom.

\subsection{Scalar waves}

Following Sec.~\ref{SecTen}, where the time-domain solutions 
were derived for a circular orbit with constant angular frequency $\omega$, we take the relative separation ${\bm r}$ 
in the $x$-$y$ plane as given by Eq.~(\ref{vbr}). 
We also choose the line-of-sight unit vector 
$\hat{\bm n}$ as defined in Eq.~(\ref{hatndef}). 
The leading time-domain solution for the scalar field 
$\varphi$ in the far zone is given by
Eq.~(\ref{phiraddetail}), namely,
\begin{equation}
\varphi(t)
=-\frac{1}{4\pi M_{\rm Pl} D} \hat{\bm n} \cdot
\dot{{\bm{\mathcal D}}}_{\phi}(t_r)\,,
\label{varphitime}
\end{equation}
where $t_r=t-D$ is the retarded time. Since the scalar dipole moment is
given by Eq.~(\ref{Dphi}), we obtain
\begin{equation}
\dot{\vb*{\mathcal D}}_{\phi}
=m_{r}\,\Delta\alpha\,r\omega(-\sin\Phi,\cos\Phi,0)\,.
\label{DphidotVIIa}
\end{equation}
Substituting this expression into 
Eq.~(\ref{varphitime}), we obtain the scalar 
wave in the time domain as
\begin{align}
\varphi(t)
&=
\frac{m_{r} \Delta\alpha}{4\pi M_{\rm Pl}D}\,
r\omega \sin\iota \,\sin\Phi \nonumber \\
&=
\frac{\eta M}{4\pi M_{\rm Pl}D}\,\Delta\alpha\,
(G_{\rm eff} M\omega)^{1/3}\sin\iota\,\sin\Phi\,,
\label{phiradtimeVII}
\end{align}
where we used $m_{r}=\eta M$ and
$v=r\omega=(G_{\rm eff} M\omega)^{1/3}$. 
Thus, the radiative scalar field oscillates at the 
first harmonic of the orbital phase, in contrast to
the leading tensor waveform, which oscillates at $2\Phi$.

We derive the frequency-domain waveform of the scalar mode by including
the radiation-reaction effect on the orbital evolution. We write
Eq.~\eqref{phiradtimeVII} as
\begin{equation}
\varphi(t)=\mathcal{A}_s(t)e^{-i\Phi(t)}
+\mathcal{A}_s^*(t)e^{i\Phi(t)},
\label{hbcomplexVII}
\end{equation}
where
\begin{equation}
\mathcal{A}_s(t)=
\frac{i\eta M}{8\pi M_{\rm Pl}D}\,\Delta\alpha\,
(G_{\rm eff} M\omega)^{1/3}\sin\iota\,.
\label{AbVII}
\end{equation}
Applying the SPA with the same Fourier convention as in
Sec.~\ref{Fouriersec}, the Fourier transform
$\tilde{\varphi}(f)=\int_{-\infty}^{\infty}{\rm d}t\,
\varphi(t)e^{-2\pi i f t}$ can be expressed as
\begin{equation}
\tilde{\varphi}(f)
\simeq
\mathcal{A}_s^*(\tilde{t}_f)
\sqrt{\frac{2\pi}{\dot{\omega}(\tilde{t}_f)}}
\,e^{-i\Psi_s(f)},
\label{hbtildeVII0}
\end{equation}
where the stationary point $\tilde{t}_f$ satisfies
\begin{equation}
2\pi f=\dot{\Phi}(\tilde{t}_f)=\omega(\tilde{t}_f)\,,
\label{phasecon}
\end{equation}
and the phase is given by
\begin{equation}
\Psi_s(f)=2\pi f \tilde{t}_f-\Phi(\tilde{t}_f)-\frac{\pi}{4}\,.
\label{Psis}
\end{equation}
Defining
\begin{equation}
\tilde{u}_f=
[G_{\rm eff}M \omega(\tilde{t}_f)]^{1/3}
=(2\pi G_{\rm eff}M f)^{1/3},
\end{equation}
and using Eqs.~\eqref{tcminustv} and \eqref{Phiv},
Eq.~\eqref{Psis} can be expressed as
\begin{align}
\Psi_s(f)
&=
2\pi f t_c-\Phi_c-\frac{\pi}{4}
+\frac{3}{256\eta}\frac{G_{\rm eff}}{G_N}
\tilde{u}_f^{-5}
\left(1-\frac{4}{7}b \tilde{u}_f^{-2}\right).
\label{Psisexplicit}
\end{align}
This shows that the leading term in $\Psi_s(f)$ scales as $f^{-5/3}$,
whereas the dipole correction proportional to $b$ scales as
$f^{-7/3}$. Using Eq.~\eqref{omegadot} with
$\omega(\tilde{t}_f)=2\pi f$, and expanding the amplitude to linear order
in $b\tilde{u}_f^{-2}$, we obtain
\begin{equation}
\tilde{\varphi} (f)
=\tilde{\varphi}_{0}\,f^{-3/2}
\left(1-\frac12 b \tilde{u}_f^{-2}\right)
e^{-i\Psi_s(f)}\,,
\label{hbtildeVII}
\end{equation}
where
\begin{equation}
\tilde{\varphi}_{0}
=-i\,\frac{\sqrt{15 \eta M}}{96\pi^{3/2} D}
\Delta\alpha \sin\iota\,.
\label{hbampVII}
\end{equation}
Thus, the leading-order amplitude of $\tilde{\varphi}(f)$ scales as
$f^{-3/2}$, reflecting the dipolar origin of the scalar radiation.

For scalar waves propagating on a cosmological background, the waveform
measured by the observer at frequency $f_0$ is obtained from
Eq.~\eqref{hbtildeVII}, together with Eqs.~\eqref{Psisexplicit} and
\eqref{hbampVII}, by making the replacements
\begin{equation}
f\to f_0,\qquad
M\to {\cal M}=(1+z)M,\qquad
D\to d_L(z),\qquad
t_c\to t_{c,0}\,.
\label{repla}
\end{equation}
Correspondingly, the velocity variable $\tilde{u}_f$ 
is replaced by $\tilde{u}=(2\pi G_{\rm eff}{\cal M} f_0)^{1/3}$.

\subsection{Vector waves}

We next discuss the radiation emitted by the vector field $A_{\mu}$.
The corresponding dipole moment is given by Eq.~(\ref{DAdetail2}).
Taking its time derivative for the circular binary separation
${\bm r}$ in Eq.~(\ref{vbr}), we obtain
\begin{equation}
\dot{\vb*{\mathcal D}}_e
=
\sqrt{4\pi\mu_\infty G_N}\,m_{r}\,
\Delta\sigma\,r\omega(-\sin\Phi,\cos\Phi,0)\,.
\label{DAdotVIIa}
\end{equation}
Substituting this expression into Eqs.~\eqref{A0so0} and \eqref{Aiso},
the radiative vector potential is given by
\begin{align}
A^0(t)
&=
\frac{1}{D}\sqrt{\frac{G_N}{4\pi\mu_\infty}}\,
m_{r}\,\Delta\sigma\,r\omega\sin\iota\,\sin\Phi\,,
\label{A0radVIIa}\\
A^i(t)
&=
-\frac{1}{D}\sqrt{\frac{G_N}{4\pi\mu_\infty}}\,
m_{r}\,\Delta\sigma\,r\omega\,(-\sin\Phi,\cos\Phi,0)\,.
\label{AiradVIIa}
\end{align}
We introduce an orthonormal basis on the plane transverse to
$\hat{\bm n}$,
\begin{equation}
{\bm e}_{\theta}=(\cos\iota,0,-\sin\iota),
\qquad
{\bm e}_{\varphi}=(0,1,0).
\end{equation}
Projecting the spatial vector potential $A^i$ in Eq.~\eqref{AiradVIIa}
onto these transverse basis vectors, and using $m_{r}=\eta M$ and
$v=r\omega=(G_{\rm eff} M\omega)^{1/3}$, we obtain two independent
transverse vector amplitudes,
\begin{align}
A_{\theta}
&\equiv \vb*{e}_{\theta}\cdot\vb*{A}
=
\frac{1}{D}\sqrt{\frac{G_N}{4\pi\mu_\infty}}\,
\eta M \Delta\sigma\,(G_{\rm eff} M\omega)^{1/3}
\cos\iota\sin\Phi\,,
\label{AthetaVII}\\
A_{\varphi}
&\equiv \vb*{e}_{\varphi}\cdot\vb*{A}
=
-\frac{1}{D}\sqrt{\frac{G_N}{4\pi\mu_\infty}}\,
\eta M \Delta\sigma\,(G_{\rm eff} M\omega)^{1/3}
\cos\Phi\,.
\label{AphiVII}
\end{align}
These quantities represent the two transverse polarization components
of the massless vector field. They oscillate at the first harmonic of
the orbital phase. The temporal component $A^0$ is not an independent
radiative degree of freedom, but is fixed by the constraint equation.
Hence, the propagating degrees of freedom are fully described by the
transverse components $A_{\theta}$ and $A_{\varphi}$.

In terms of the complex representation, Eqs.~(\ref{AthetaVII}) and
(\ref{AphiVII}) can be written as
\begin{equation}
A_{\theta}(t)
=
\mathcal{A}_{\theta}(t)e^{-i\Phi(t)}
+\mathcal{A}_{\theta}^*(t)e^{i\Phi(t)},
\end{equation}
where
\begin{equation}
\mathcal{A}_{\theta}(t)
=
\frac{i}{2D}\sqrt{\frac{G_N}{4\pi\mu_\infty}}\,
\eta M \Delta\sigma\,( G_{\rm eff} M\omega)^{1/3}\cos\iota\,,
\label{AthetaAmpVII}
\end{equation}
and
\begin{equation}
A_{\varphi}(t)
=
\mathcal{A}_{\varphi}(t)e^{-i\Phi(t)}
+\mathcal{A}_{\varphi}^*(t)e^{i\Phi(t)},
\end{equation}
with
\begin{equation}
\mathcal{A}_{\varphi}(t)
=
-\frac{1}{2D}\sqrt{\frac{G_N}{4\pi\mu_\infty}}\,
\eta M \Delta\sigma\,( G_{\rm eff} M\omega)^{1/3}\,.
\label{AphiAmpVII}
\end{equation}
Applying the SPA with the same stationary condition~\eqref{phasecon} as
in the scalar case, the frequency-domain waveforms of the vector modes
are given by
\begin{align}
\tilde A_{\theta}(f)
&=
\tilde A_{\theta,0}\,f^{-3/2}
\left(1-\frac12 b \tilde{u}_f^{-2}\right)e^{-i\Psi_s(f)}\,,
\label{AthetatildeVII}\\
\tilde A_{\varphi}(f)
&=
\tilde A_{\varphi,0}\,f^{-3/2}
\left(1-\frac12 b \tilde{u}_f^{-2}\right)e^{-i\Psi_s(f)}\,,
\label{AphitildeVII}
\end{align}
where $\Psi_s$ is defined in Eq.~(\ref{Psis}) and can be written
explicitly as Eq.~(\ref{Psisexplicit}), 
and the frequency-independent prefactors are
\begin{align}
\tilde A_{\theta,0}
&=
-i\,\frac{\sqrt{30\eta M}}{192\pi^{3/2}\sqrt{\mu_\infty}\,D}
\Delta\sigma\cos\iota\,,
\label{AthetaampVII}\\
\tilde A_{\varphi,0}
&=
-\frac{\sqrt{30\eta M}}{192\pi^{3/2}\sqrt{\mu_\infty}\,D}
\Delta\sigma\,.
\label{AphiampVII}
\end{align}
Thus, when vector radiation is described in terms of the vector potential
itself, its frequency-domain amplitude scales as $f^{-3/2}$, in close
analogy with the scalar waveform. If one instead considers the radiative
electric field~\eqref{Ejdef}, the additional time derivative increases
the frequency dependence by one power, so that the corresponding Fourier
amplitudes scale as $f^{-1/2}$.

For vector waves propagating on a cosmological background, the waveforms
measured by an observer at frequency $f_0$ can be obtained from
Eqs.~(\ref{AthetatildeVII}) and (\ref{AphitildeVII}) by applying the
same replacements as those given in Eq.~(\ref{repla}).

To summarize, ESM theories give rise to radiative 
scalar and vector fields whose leading amplitudes are 
of dipolar origin and are therefore proportional to the first harmonic of the orbital phase. In the absence
of nonminimal couplings to gravity, however, 
the metric waveform still contains only the tensor 
polarizations discussed in Sec.~\ref{Fouriersec}.
The scalar and vector sectors nevertheless leave 
observable imprints through the dipole-flux 
correction encoded in the combination
$(\Delta\alpha)^2+(\Delta\sigma)^2$.

\bibliographystyle{mybibstyle}
\bibliography{bib}

\end{document}